%% file: QuatZip.tex
\pdfoutput=1
\newcommand{\BlindSubmission}{0}
\newcommand{\ArxivVersion}{}
\newcommand{\SupplementalName}{appendix}
\makeatletter
\def\input@path{{./template/}}
\makeatother
\documentclass{egpubl}
\usepackage{eg2026}
\ConferenceSubmission   % uncomment for Conference submission    
\CGFStandardLicense
\usepackage[T1]{fontenc}
\usepackage{dfadobe}  
\usepackage{cite}
\BibtexOrBiblatex
\electronicVersion
\PrintedOrElectronic
\usepackage[pdftex]{graphicx}
\usepackage{egweblnk} 

\input{common/packages.tex}
\usepackage{amssymb}
\input{common/commands.tex}

\graphicspath{{./}, {../figs}, {./template/}} 
\ifthenelse{\equal{\BlindSubmission}{1}}
{
}
{
    
    \author[Faber et al.]
    {\parbox{\textwidth}{\centering   
            Carsten Faber$^1$\orcid{0009-0008-2879-3969}
            Bastian Kuth$^2$\orcid{0000-0001-9473-8847}
            Rinaldo Schuster$^1$
            Daniel Pauls$^1$
            Quirin Meyer$^2$\orcid{0000-0001-7073-442X}            
            }
            \\
            {
                \parbox{\textwidth}
                {
                    \centering $^1$Coburg University of Applied Sciences and Arts, Germany $\quad ^2$AMD, Germany
                }
            }
    }
}
\glsdisablehyper
\makeglossaries
\input{lists/acronyms.tex}

\title{Compression of Uniformly Distributed Unit Quaternions}
\begin{document}
\teaser
{
    \input{./secs/A0teaser.tex}
}
\maketitle
\begin{abstract}
\input{secs/A0abstract.tex}
%  ACM CCS 2012 (see https://www.acm.org/publications/class-2012)
%  Mapped from the submission categories: Rigging/Skinning, GPUs/Graphics Hardware, Geometry, Mathematical Analysis,
%  Optimization, Compression, Real-Time Rendering.
%Use http://dl.acm.org/ccs.cfm to generate the file.

\end{abstract}
\input{secs/structure.tex}

\input{secs/C0ack.tex} 
\bibliographystyle{template/eg-alpha-doi}
\bibliography{common/bibliography}
\appendix
\input{secs/S0proofs.tex}
\input{secs/S1symbols.tex}

\end{document}

%% file: common/packages.tex
\usepackage{tikz}
\usetikzlibrary{positioning}
\usetikzlibrary{calc} 
\usepackage{placeins}
\usepackage{siunitx}
\usepackage{enumitem}
\usepackage{subcaption}
\usepackage{multirow}
\usepackage{mathtools}
\mathtoolsset{showonlyrefs=true}
\usepackage{makecell}
\usepackage{array}
\usepackage{lipsum}
\usepackage{amsmath}
\usepackage{amssymb}
\usepackage{pifont}
\usepackage{ifthen} 
\usepackage{hyperref}
\usepackage[T1]{fontenc}
\usepackage[utf8]{inputenc}
\usepackage{layout}
\usepackage{comment}
\usepackage{listings}
\usepackage{xcolor}
\usepackage{accsupp}  % ActualText for \FLIP (common/commands.tex).
\usepackage[outline]{contour}  % Outlined labels on images (teaser): \contour{black}{text}.
\contourlength{0.35pt}
\usepackage{nicefrac}
\usepackage[inkscapepath=build/svg-inkscape]{svg}
\ifdefined\ArxivVersion\svgsetup{inkscape=false}\fi
\usepackage[toc, nogroupskip, nonumberlist, nopostdot, acronyms, shortcuts, translate=babel]{glossaries}
\usepackage{tabularray}
\makeatletter
\@ifpackageloaded{booktabs}{}{
  \UseTblrLibrary{booktabs}
}
\makeatother
\UseTblrLibrary{diagbox}

\usepackage[capitalise,noabbrev]{cleveref}
\crefname{section}{Sec.}{Secs.}
\crefname{subsection}{Sec.}{Secs.}
\crefname{subsubsection}{Sec.}{Secs.}
\crefname{table}{Tab.}{Tabs.}
\crefname{figure}{Figure}{Figures}
\crefname{equation}{Eq.}{Eqs.}
\Crefname{section}{Section}{Sections}
\Crefname{subsection}{Section}{Sections}
\Crefname{subsubsection}{Section}{Sections}
\Crefname{table}{Table}{Tables}
\Crefname{figure}{Figure}{Figures}
\Crefname{equation}{Equation}{Equations}
\creflabelformat{equation}{#2(#1)#3}

\makeatletter
\renewcommand*\MT@newlabel[1]{\global\@namedef{MT_r_#1}{}\global\@namedef{MT_r_{#1}}{}}
\makeatother
\NewCommandCopy{\crefOriginal}{\cref}
\NewCommandCopy{\CrefOriginal}{\Cref}
\RenewDocumentCommand{\cref}{s m}{\noeqref{#2}\IfBooleanTF{#1}{\crefOriginal*{#2}}{\crefOriginal{#2}}}
\RenewDocumentCommand{\Cref}{s m}{\noeqref{#2}\IfBooleanTF{#1}{\CrefOriginal*{#2}}{\CrefOriginal{#2}}}

%% file: common/commands.tex
\newcommand{\mat}[1]{\mathbf{#1}}

\newcommand{\vect}[1]{\mathbf{#1}}      % vector (bold)
\newcommand{\uvec}[1]{\hat{\vect{#1}}}  % unit vector (bold + hat)
\newcommand{\ucomp}[1]{\hat{#1}}        % component of a unit vector (hat, no bold -- it's a scalar)
\newcommand{\quant}[1]{\underline{#1}}  % quantized value (scalar or vector)
\newcommand{\aprx}[1]{\tilde{#1}}       % approximated value (scalar or vector)

\newcommand{\lOne}{\ell_1}
\newcommand{\lTwo}{\ell_2}
\newcommand{\lTwoI}{\ell_{2\mathrm{I}}}
\newcommand{\lInf}{\ell_\infty}
\newcommand{\lInfT}{\ell_{\infty}\mathrm{t}}
\newcommand{\lInfTan}{\ell_{\infty}\mathrm{tan}}
\newcommand{\lInfCub}{\ell_{\infty}\mathrm{cub}}

\newcommand{\FLIP}{\text{\BeginAccSupp{method=escape,ActualText=FLIP}\reflectbox{F}LIP\EndAccSupp{}}}

\newcommand{\amdGpu}{AMD Radeon\texttrademark{} RX~7900~XTX}
\newcommand{\nvidiaGpu}{NVIDIA\textsuperscript{\textregistered} GeForce RTX\texttrademark{}~4070~Ti}

\providecommand{\SupplementalName}{supplemental material}
\ifdefined\ArxivVersion\input{common/arxiv.tex}\fi

\definecolor{todocolor}{RGB}{220,0,120}

%% file: lists/acronyms.tex
\newcommand{\acro}[2]{\newacronym{#1}{#1}{#2}}
\acro{GPU}{graphics processing unit}
\acro{CPU}{central processing unit}
\acro{ALU}{arithmetic and logical unit}
\acro{API}{application programming interface}
\acro{3D}{three-dimensional}
\acro{2D}{two-dimensional}
\acro{1D}{one-dimensional}
\acro{bpt}{bits per triangle}
\acro{bpv}{bits per vertex}
\acro{LUT}{lookup table}
\acro{SIMD}{single instruction, multiple data}
\acro{AABB}{axis-aligned bounding box}
\acro{GLSL}{OpenGL Shading Language}
\acro{rANS}{range asymmetrical numeral systems}
\acro{POT}{power-of-two}
\acro{LBS}{linear blend skinning}
\acro{DQS}{dual quaternion skinning}
\acro{LMS}{log-matrix skinning}
\acro{SBS}{spherical blend skinning}
\acro{CRS}{centre of rotation-optimized skinning}
\acro{i.i.d.}{independent and identically distributed}
\acro{OSS}{Optimal Simplex Sampling}
\acro{LP}{linear program}
\acro{ILP}{integer linear program}
\acro{MILP}{mixed integer linear program}
\acro{GTS}{generalized triangle strip}
\acro{ATS}{alternating triangle strip}
\acro{CLZ}{count leading zeros}
\acro{TEA}{Tiny Encryption Algorithm}
\acro{ETA}{Enhanced Tunneling Algorithm}
\acro{LW}{Laced Wires}
\acro{Gtps}{Giga triangles per second}

%% file: secs/A0teaser.tex
% Teaser layout (work in progress): animation and tangent frames (2.57 in, as narrow as the lions and their error
% scale allow; the scale serves both) in the top row, Gaussian splatting over the full width under it. The panels
% come from doc/TeaserFigure.
% 2.05 in for the top row is the most that keeps teaser and abstract (with the two-line CCS concepts) on the
% first page of the named version; 2.1 in runs into the footer.
\centering
\setlength{\fboxsep}{0pt}%
\makebox[\textwidth][c]{%
\includesvg[inkscapearea=page,width=\dimexpr\textwidth-2.61in\relax]{figs/teaser_animation_dragon}%
\hspace{0.04in}%
\includesvg[inkscapearea=page,width=2.57in]{figs/teaser_tangent_lion}}\\[0.04in]
\makebox[\textwidth][c]{%
\includesvg[inkscapearea=page,width=\textwidth]{figs/teaser_gaussian_bike}}
\caption[The teaser image.]{%
    Three applications for our unit quaternions compressing method $\lInfCub$.
    \emph{Skeletal Animation} (top left):
    with uncompressed rotations (upper row) and the error of $\lInfCub$ with \num{48} bits per rotation key (lower row), grey where exact, else colored by ${e=\max(|\Delta R|,|\Delta G|,|\Delta B|)}$ in 8-bit levels.
    \emph{Normal Mapping} (top right): 
    the reference and $\lInfCub$ with \num{32} bits per tangent frame, split by the diagonal, and the error $e$ of $\lInfCub$.
    \emph{3D Gaussian Splatting} (bottom):
    the reference, $\lInfCub$ with \num{30} bits per quaternion rotation, and its \FLIP{} error.
    % The dragon and the bicycle are credited in the acknowledgments (secs/C0ack.tex) and the results figures.
}\label{fig:teaser}

%\num{32} bits encode one tangent frame for normal mapping:

%rotations of Gaussian splats with \num{30} bits each: 

%% file: secs/A0abstract.tex
Unit quaternions are widely used in computer graphics, making their efficient compression critical for reducing memory consumption and bandwidth requirements.
In this paper, we present a novel method for encoding uniformly distributed unit quaternions.
Our thorough analysis of several quaternion parameterizations compares their theoretical error properties, quantization behavior, computational cost, and memory efficiency and provides both an analytical and empirical characterization.
We show that an infinity-norm parameterization with our optimized warp function consistently offers the smallest maximum error and the best storage efficiency, at a decompression cost equal to or lower than that of parameterization methods found in practice.
We demonstrate the practical benefits of the proposed approach in several graphics applications, including tangent-space normal mapping, skeletal animation with blend weights, and Gaussian splatting.

%% file: secs/structure.tex
% A0abstract.tex is \input'ed in template/main.tex.
% A0teaser.tex is \input'ed in template/main.tex.
\input{./secs/B1intro.tex}
\input{./secs/B2background.tex}

\input{./secs/B3main.tex}
\input{./secs/B4warpfunctions.tex}
\input{./secs/B5results.tex}
\input{./secs/B6conclusion.tex}

%% file: secs/B1intro.tex
\section{Introduction}\label{Sec:Introduction}
Unit quaternions are a standard representation of 3D rotations in computer graphics.
They are compact, avoid gimbal lock, interpolate smoothly, and are cheap to compose and to convert into rotation matrices.
Consequently, they appear throughout real-time and offline pipelines:
skeletal animation stores one quaternion per joint and key frame,
the tangent frames of mesh vertices can be stored as quaternions for normal mapping,
and 3D Gaussian splatting stores the orientation of millions of anisotropic Gaussians as quaternions.

Since memory size and bandwidth, unlike computational speed, increasingly limit performance, every byte saved per rotation reduces storage, transfer, and cache pressure.
Hence, compression of unit quaternions has been widely adopted, for example, 
in animation libraries~\cite{ozzAnimation,frechette2020acl},
asset formats~\cite{Kapoulkine20MeshoptGLTF},
tangent frame encodings~\cite{Frey11SSD,Kapoulkine26QTF},
and Gaussian splatting formats~\cite{NianticSPZ,PlayCanvasSOG}.

A good quaternion compression method has to meet several goals.
It should be \emph{compact} and use as few bits as possible for a given accuracy, have a \emph{small maximum and mean error}, 
and \emph{decompress fast}, ideally with a few arithmetic instructions and without lookup tables.
Compression, which usually runs once per asset, should be fast enough for asset pipelines.

Besides using no compression at all~\cite{Kerbl233DG}, different quaternion compression methods are used, such as
dropping one component~\cite{frechette2020acl}, dropping the largest component~\cite{ozzAnimation, Kapoulkine26QTF}, projecting onto the hyper-cube~\cite{ISO05ITC}, projecting onto the hyper-octahedron~\cite{Kuth25RTG, Kapoulkine26QTF}, or mapping them into the unit ball~\cite{Reynolds17QQ,rijsdijk2026GPS}.
While compression methods for 3D unit vectors have been compared extensively~\cite{Meyer10OFN,Meyer12RTG}, 
only empirical results exist for unit quaternions~\cite{Reynolds17QQ,Kapoulkine26QTF}.
In this paper, we fill this gap and make the following contributions.
\begin{itemize}
    \item In \cref{Sec:MainPart}, we study commonly used unit vector parameterizations and their maximum quantization errors.
    \item We present a mixed-radix packing for unit quaternions that stores the chart index and all quantized parameters as a single integer code.
    Unlike separate bit fields, it uses almost all states of a code word of any size while giving every parameter the same precision.
    \item In \cref{Sec:WarpFunctions}, we introduce the $\lInfCub$-parameterization, which uses a cubic warp function whose single parameter we optimize to minimize the maximum error.
    It achieves the lowest maximum error of all compared parameterizations.
    \item In \cref{Sec:Evaluation}, we evaluate $\lInfCub$ in skeletal animation, normal mapping, and Gaussian splatting, and show that it is more accurate than the codec in use in each of these domains at the same storage cost, at an equal or lower decoding cost.
\end{itemize}

Our $\lInfCub$ is a drop-in replacement for existing quaternion codecs such as shown in our teaser~\cref{fig:teaser}, yielding a smaller maximum error at the same bit budget and without additional decoding cost.

Common to parallel decompression codecs~\cite{Kuth25RMD}, we limit the scope of the paper to fixed-rate codes that decode each quaternion independently and thus allow random access.
Further, we assume that the unit quaternions are uniformly distributed.

%% file: secs/B2background.tex
\section{Previous Work}\label{sec:PreviousWork}
Shoemake~\cite{Shoemake85ARQ} introduced unit quaternions to computer graphics to represent and interpolate rotations.
The smallest-three method~\cite{Fiedler15SC}, called $\lTwoI$ in this paper, is used, for example, by the animation library ozz-animation~\cite{ozzAnimation}, the Khronos-ratified glTF extension \texttt{EXT\_meshopt\_compression}~\cite{Kapoulkine20MeshoptGLTF}, and the Gaussian splatting formats SPZ~\cite{NianticSPZ} and SOG~\cite{PlayCanvasSOG}.
The Animation Compression Library~\cite{frechette2017gdc, frechette2020acl} and the classic codecs of Unreal Engine~4~\cite{Frechette17UE4} always drop the same component~($\lTwo$).
The Gaussian splatting renderer of Rijsdijk et~al.~\cite{rijsdijk2026GPS} quantizes quaternions with a method due to Reynolds~\cite{Reynolds17QQ}:
There, a unit quaternion is first mapped to the unit ball and then compressed.
Alexa~\cite{Alexa22SFS} maps indices to orientations that are distributed almost uniformly.
While decompression is $\mathcal{O}(1)$, Alexa provides no inverse mapping, so compression requires a closest-point search over all orientations.
This is feasible with a precomputed table for small bit budgets, but not for large ones:
for 32-bit indices, a table of $2^{32}$ quaternions with four 32-bit floats each requires \qty{64}{\gibi\byte}.

Quaternions can be seen as a 4D generalization of 3D unit vectors.
For those, many compression schemes exist~\cite{Deering95GC, Griffith07FNV, Keinert15SFM, Rousseau20UUV}.
Octahedral normal vectors~\cite{Meyer10OFN} have been used in games~\cite{Geffroy20DOOM, karis2021nanite} and movies~\cite{Eisenacher13SDS, Christensen25RMX}.
As with unit quaternions, range quantization improves the error of unit vectors~\cite{Meyer10OFN, Kaplanyan10CE3, Meyer12RTG}.
There exist generic methods for attribute quantization~\cite{Calver2002VDS, Purnomo05HVC}, as well as special variants for positions~\cite{Lee10C3M, Meyer11ALP, Limper13PBR, Kwan18PVD} and blend attributes~\cite{Kuth21VBA, Peters22PCV}.

CryEngine stores tangent frames as quaternions with four 16-bit components, called QTangents~\cite{Frey11SSD}.
They encode the handedness in the sign of the fourth component and bias that component away from zero, so that its sign survives quantization.
DOOM Eternal fits a tangent frame into only three bytes, two for an octahedral normal vector and one for the angle of the tangent around the normal, without handedness~\cite{Geffroy20DOOM}.

MPEG-4~\cite[Sec.~8.3.3]{ISO05ITC} quantizes unit quaternions with the tangent-warped $\lInf$-parameterization, called $\lInfTan$ in this paper.
Lambers~\cite{Lambers20SCM} surveys warps that map the faces of a cube more uniformly onto the sphere.
Zucker and Higashi~\cite{Zucker18CSP} optimize odd polynomials, a family that includes our cubic warp (\cref{Sec:WarpFunctions}), to minimize the area distortion on the 2-sphere.
In contrast, we minimize the maximum angular error of quantized unit quaternions on the 3-sphere.

Mixed-radix vector compression of \cref{Sec:PackAndUnpack} has been proposed for vertex-blend attribute quantization~\cite{Kuth21VBA}.
It resembles asymmetric numeral systems with a uniform distribution~\cite{Duda14ANS}.

%% file: secs/B3main.tex
\section{Parameterization Methods for Unit Vectors}\label{Sec:MainPart}
Before we discuss quaternion compression methods in \cref{Sec:L2,Sec:Parameterizations}, \cref{Sec:Notation} introduces the notation.
We present our quantization scheme (\cref{Sec:Quantization}), propose mixed-radix packing (\cref{Sec:PackAndUnpack}), and contribute a thorough error analysis (\cref{Sec:ErrorAnalysis}).

\subsection{Notation}\label{Sec:Notation}
We mark vectors with boldface, e.g., $\vect{x}$, and additionally mark \emph{unit vectors}, where $\Vert \vect{x}\Vert_2=1$ holds, with a hat, e.g., $\uvec{x}=[\ucomp{x}_1,\dots,\ucomp{x}_4]^\top\in\mathbb{R}^4$.
$\mathrm{norm}_k\left(\vect{x}\right):=\vect{x} / \Vert \vect{x} \Vert_k$ normalizes a vector by its $k$-norm.
A scalar component of a unit vector, e.g., $\ucomp{x}_1$, keeps the hat but not the boldface.
$\vect{x}_{1:3}=[x_1,x_2,x_3]^\top$ denotes the first three components of a vector.
Quantized scalars or vectors are underlined, e.g., $\quant{u}$, $\quant{\vect{u}}$.

We treat \emph{unit quaternions} as 4D unit vectors~\cite{Shoemake85ARQ} and write them as $\uvec{x}$ with $\Vert\uvec{x}\Vert_2^2=\sum_{i=1}^4 \ucomp{x}_i^2=1$.
All unit quaternions form the boundary surface of a 4D hyper-sphere called $S^{3}$.

We use differential-geometry terminology~\cite{Crane2022DDG}.
$S^{3}$ is a 3D manifold embedded in 4D space.
Therefore, a few mappings -- called \emph{charts} --  that map $\mathbb{R}^{3} \rightarrow S^{3}\subset\mathbb{R}^4$ are sufficient to represent a unit vector.
We call a set of charts that cover $S^{3}$ entirely a \emph{parameterization} $\vect{f}(\vect{u}; \vect{c})$ -- an \emph{atlas} in differential-geometry terminology -- where $\vect{u}\in\mathbb{R}^{3}$ is the \emph{(domain) parameter} and $\vect{c}$ is the \emph{chart index}.
The combination $\left[\vect{u}; \vect{c}\right]^\top$ is called \emph{atlas coordinates}.
The parameterizations used in this work require one chart index $c\in\mathbb{N}$ or two chart indices $\vect{c}\in\mathbb{N}^2$.
The number of charts for $c_i$ is $C_i$.
This way, we can compress the four values of a unit quaternion to three values and a memory-efficient \emph{chart index} $c$.
Here, we do not require more than eight charts, so $\vect{c}$ uses three bits at most.

\subsection{A Simple Parameterization}\label{Sec:L2}
There are several methods to compress unit vectors to atlas coordinates.
They compress a direction vector $\vect{x}$, with $\vect{x}\neq\vect{0}$, and decompress to a unit vector $\uvec{x}$.
Because we focus on real-time graphics, we pick parameterizations that are quickly evaluated and avoid expensive functions such as trigonometric functions.

As an introductory example, we use 2D unit vectors $\uvec{x}$.
We can compute one component from the other using two charts.
Chart $0$ covers the half circle for which $\ucomp{x}_2\geq 0$:
\begin{equation}
\vect{f}_{\lTwo}(u; 0)=
\begin{bmatrix}
    \ucomp{x}_1\\
    \ucomp{x}_2
\end{bmatrix}=
\begin{bmatrix}
    u\\+\sqrt{1-u^2}
\end{bmatrix}
\end{equation}
and chart $1$ the half circle for which $\ucomp{x}_2 < 0$:
\begin{equation}
\vect{f}_{\lTwo}(u; 1)=
\begin{bmatrix}
    \ucomp{x}_1\\
    \ucomp{x}_2
\end{bmatrix}=
\begin{bmatrix}
    u\\-\sqrt{1-u^2}
\end{bmatrix}.
\end{equation}
For unit quaternion compression, the $\lTwo$-parameterization drops one component~\cite{frechette2020acl} and decompresses to
\begin{equation}
\vect{f}_{\lTwo}(\vect{u}; c)=
    \uvec{x} =
\begin{bmatrix}
    \vect{u}\\
    (-1)^c \cdot {\sqrt{1-\Vert\vect{u}\Vert_2^2}}
\end{bmatrix}.
\end{equation}
A subscript of $\vect{f}$ indicates the parameterization.
We use $\vect{f}^{-1}$ for compression. 
It maps a unit vector $\uvec{x}$ to the parameters $\vect{u}$ and the chart index $c$.
For the $\lTwo$-parameterization, that is
\begin{equation}
        \vect{f}_{\lTwo}^{-1}\left(\uvec{x}\right) =
        \begin{bmatrix}
            \vect{u}\\
            c
        \end{bmatrix}
        =
        \begin{bmatrix}
            \uvec{x}_{1:3}\\
            \mathrm{neg}(\ucomp{x}_4)
        \end{bmatrix},
\end{equation}
where $\mathrm{neg}(x)=1$ if $x<0$ and $\mathrm{neg}(x)=0$ otherwise.
To avoid negative square-root arguments, the decoder projects $\vect{u}$ onto the unit ball, $\Vert\vect{u}\Vert_2\leq1$.

\subsection{Quantization}\label{Sec:Quantization}
While the chart indices $\vect{c}$ are discrete, the parameters $u_i$ are continuous and must be quantized.
We choose to uniformly quantize $u_i$ with a fixed-point representation.
We refrain from using floats, because neither their logarithmic distribution nor their vast range offers any benefit.
Additionally, floats have fixed bit sizes of 16, 32, or 64 bits, which is impractical for fine-tuning compression ratios against quality.
In contrast, with fixed-point, we can configure the number of bits or even the number of discrete states $U$ for $u_i$.

All parameterizations below use parameters $u\in[-1,1]$.
To quantize $u$ with $U\in\mathbb{N}$ discrete values, we use
\begin{equation}
 \quant{u}={q}\left(u\right)=
 \min\left(
 \left\lfloor \frac{u+1}{\varepsilon} \right\rfloor,U-1\right)
\end{equation}
and dequantize it with
\begin{equation}
    \aprx{u} = {q}^{\sim 1}\left(\quant{u}\right) = 
    \left(\quant{u} + \frac{1}{2}\right) \cdot \varepsilon  - 1 ,
\end{equation}
where $\varepsilon=\frac{2}{U}$ is the \emph{sample spacing}.
The $\sim 1$ indicates that dequantization is a lossy inverse of ${q}$.
Note that quantizing and dequantizing together round $u$ to the nearest sample.

For \textbf{domain quantization}, we map an \emph{input unit vector} $\uvec{x}$ to atlas coordinates $\left[\vect{u};\vect{c}\right]^\top$ and quantize them to $\left[\quant{\vect{u}};\vect{c}\right]^\top$.
When reversing the process, we obtain a \emph{reconstructed vector} $\aprx{\uvec{x}}$.
Since quantization introduces errors, $\aprx{\uvec{x}}$ and $\uvec{x}$ usually differ.
To assess the \textbf{quantization error}, we must quantify how much the two vectors differ in the directions they point to.
Therefore, we use the geodesic distance, i.e., the angle between $\aprx{\uvec{x}}$ and $\uvec{x}$:
\begin{equation}
    \angle\left(\uvec{x},\aprx{\uvec{x}}\right)=\arccos
        {
            \frac
            {\aprx{\uvec{x}}^\top \uvec{x}}
            {\Vert\aprx{\uvec{x}}\Vert_2 \Vert\uvec{x}\Vert_2}
        }.
        \label{Eq:QuantizationError}
\end{equation}

Domain quantization does not produce the smallest possible quantization error, because it quantizes in the domain $\vect{u}$ rather than in the range of $\vect{f}$.
We obtain a lower error using \textbf{range quantization}:
There, we map the unit vector to the domain $\left[\vect{u};\vect{c}\right]^\top = \vect{f}^{-1}(\uvec{x})$.
Then $\vect{u}$ lies in a cell whose $2^3=8$ corners are samples.
Those corners are the quantization candidates.
Range quantization chooses the candidate minimizing the geodesic distance.
This is similar to how 3D unit vectors are compressed best~\cite{Meyer10OFN}.

\subsection{Mixed-Radix Packing}\label{Sec:PackAndUnpack}
We pack the quantized atlas coordinates $\left[\quant{\vect{u}};\vect{c}\right]^\top$ into a system word, typically of 16, 32, or 64 bits, but any other bit size is possible.
\input{codes/PackUnpack.tex}
In our case, the radices are constant integers and the compiler replaces divide and modulo operations with faster ones.

Mixed-radix packing is especially useful for quaternion compression:
A method with eight charts requires three bits for the chart index.
When packing a quaternion into a 32-bit word, that leaves $29/3\approx9.7$ bits for each of the three components.
Unless we favor some components over others, only $9$ bits per component are usable with bitfields, leaving $2$ bits of the word unused.
With mixed-radix packing, $29/3$ bits amount to $U=812$ states instead of $512$.
This increases the accuracy of the compressed quaternions.

\subsection{Parameterizations}\label{Sec:Parameterizations}

\noindent\textbf{$\lOne$-Parameterization}:
Instead of the $\lTwo$-norm, we can use the $\lOne$-norm~\cite{Kuth25RTG,Kapoulkine26QTF}.
First, we take an input vector $\vect{x}$ and project it onto the hyper-octahedron by dividing the components by its $\lOne$-norm.
Thus, we can compute $x_4$ from the other components, except for its sign.
Therefore, during compression, we store only its sign as the chart index $c$ and the non-redundant components in $\vect{u}$:
\begin{equation}
        \vect{f}_{\lOne}^{-1}\left(\vect{x}\right) =
        \begin{bmatrix}
            \vect{u}\\
            c
        \end{bmatrix}
        =
        \begin{bmatrix}
            \frac{\vect{x}_{1:3}} { \Vert\vect{x}\Vert_1}\\
            \mathrm{neg}(x_4)
        \end{bmatrix}.
\end{equation}
During decompression, we reverse the process by normalizing an unnormalized vector $\vect{v}$:
\begin{equation}
\vect{f}_{\lOne}(\vect{u}; c)=
    \uvec{x} =
    \mathrm{norm}_2\left(\vect{v}\right),\quad
    \vect{v} =
    \begin{bmatrix}
        \vect{u}\\
        (-1)^c \cdot \left({1 - \Vert\vect{u}\Vert_1}\right)
    \end{bmatrix}.
\end{equation}
 
\noindent\textbf{$\lTwoI$-Parameterization}:
The $\lTwo$-parameterization becomes unstable for $\ucomp{x}_4\rightarrow0$ due to the square root becoming infinitely steep there.
This can be avoided~\cite{Fiedler15SC, Kapoulkine20MeshoptGLTF, Kapoulkine26QTF} by finding the component of the largest magnitude of a unit vector $\uvec{x}$
\begin{equation}
    i_\mathrm{max}=\arg\max_{i\in\left\{1,2,3,4\right\}}\vert \ucomp{x}_i \vert,
\end{equation}
and permuting the components of $\uvec{x}$ with the permutation function $\vect{x} = \vect{P}^{-1}_{i_\mathrm{max}}(\uvec{x})$, which swizzles the components of $\uvec{x}$ such that the component of the largest magnitude becomes the last component:
\begin{equation}
    \vect{P}^{-1}_{i_\mathrm{max}}(\uvec{x})_i = \ucomp{x}_{\left[\left(i + i_\mathrm{max}-1\right)\bmod 4\right]+1}\label{Eq:Swizzle}.
\end{equation}
We encode its first three components in $\vect{u}$, $i_\mathrm{max}$ as the chart index $c_1$, and whether $\vect{P}^{-1}_{i_\mathrm{max}}(\uvec{x})_4$ is negative as the chart index $c_2$:
\begin{equation}
        \vect{f}_{\lTwoI}^{-1}\left(\uvec{x}\right) =
        \begin{bmatrix}
            \vect{u}\\
            \vect{c}
        \end{bmatrix}
        =
        \begin{bmatrix}
            \left(\vect{P}^{-1}_{i_\mathrm{max}}\uvec{x}\right)_{1:3}\\
            i_\mathrm{max}\\
            \mathrm{neg}\left(\left(\vect{P}^{-1}_{i_\mathrm{max}}\uvec{x}\right)_4\right)
        \end{bmatrix}.
\end{equation}
For decompression, $\vect{P}_{i_\mathrm{max}}$ reverts the permutation:
\begin{equation}
\vect{f}_{\lTwoI}(\vect{u}; \vect{c})=
    \uvec{x} =
    \vect{P}_{i_\mathrm{max}}\left(\vect{v}\right),\quad
    \vect{v} =
    \begin{bmatrix}
        \vect{u}\\
        (-1)^{c_2}\sqrt{1-\vect{u}^\top\vect{u}}
    \end{bmatrix}.
\end{equation}
Here, $\vect{v}$ is already of unit length, so normalizing it is unnecessary.
Since the last component of $\vect{P}^{-1}_{i_\mathrm{max}}(\uvec{x})$ has the largest magnitude, $u_i^2\leq\frac{1}{2}$, i.e., $u_i\in\left[-\frac{\sqrt{2}}{2},\frac{\sqrt{2}}{2}\right]$.
We scale $u_i$ by $\sqrt{2}$ to $[-1,1]$ before quantization and undo the scaling after dequantization.

\noindent\textbf{$\lInf$-Parameterization} is similar to the $\lTwoI$-parameterization, except that we project $\vect{x}$ onto the hyper-cube by dividing the components by the maximum norm after permutation.
Then, the last component is $\vect{P}^{-1}_{i_\mathrm{max}}(\vect{x})_4 = \pm 1$, and we store only its sign:
\begin{equation}
        \vect{f}_{\lInf}^{-1}\left(\vect{x}\right) =
        \begin{bmatrix}
            \vect{u}\\
            \vect{c}
        \end{bmatrix}
        =
        \begin{bmatrix}
            \frac{\left(\vect{P}^{-1}_{i_\mathrm{max}}\vect{x}\right)_{1:3}} { \Vert \vect{x}\Vert_\infty}\\
            i_\mathrm{max}\\
            \mathrm{neg}\left(\left(\vect{P}^{-1}_{i_\mathrm{max}}\vect{x}\right)_4\right)
        \end{bmatrix}.
\end{equation}
For decompression, we undo the permutation and normalize:
\begin{equation}
\vect{f}_{\lInf}(\vect{u}; \vect{c})=
    \uvec{x} =
    \mathrm{norm}_2\left(\vect{P}_{i_\mathrm{max}}\left(\vect{v}\right)\right),\quad
    \vect{v} =
    \begin{bmatrix}
        \vect{u}\\
        (-1)^{c_2}
    \end{bmatrix}.
\end{equation}

\noindent\textbf{$\lInf$-Warped-Parameterization}:
When warping $\vect{u}$ during the $\lInf$-parameterization, we can decrease the quantization error:
\begin{equation}
\vect{f}_{\lInfT}(\vect{u}; \vect{c})=\vect{f}_{\lInf}\left(\vect{w}\left(\vect{u}\right);\vect{c}\right),
\end{equation}
where $\vect{w}$ applies the same function $w$ to all components, so that $\vect{v}=\left[\vect{w}\left(\vect{u}\right)^\top, (-1)^{c_2}\right]^\top$.
A known good choice for $w$ that reduces the quantization errors is $w_\mathrm{tan}(x) = \tan{\left(\frac{\pi}{4}\cdot x\right)}$~\cite{ISO05ITC}.
In fact, $w_\mathrm{tan}$ is perfect for 2D.
We call the resulting parameterization $\lInfTan$.
However, the tangent is computationally expensive and not optimal for 3D and 4D, so we discuss various warp functions in \cref{Sec:WarpFunctions}.
During compression, we undo the warping after the projection:
\begin{equation}
        \vect{f}_{\lInfT}^{-1}\left(\vect{x}\right) =
        \begin{bmatrix}
            \vect{w}^{-1}\left(\vect{u}'\right)\\
            \vect{c}
        \end{bmatrix},\quad
        \begin{bmatrix}
            \vect{u}'\\
            \vect{c}
        \end{bmatrix}
        =\vect{f}_{\lInf}^{-1}\left(\vect{x}\right),
\end{equation}
requiring $w$ to be an invertible function.

\noindent\textbf{Exp-Parameterization}
first converts a quaternion into a rotation axis and angle~\cite{Reynolds17QQ}. 
Scaling the unit-length axis by half the angle gives a point inside a 3D ball with radius $\frac{\pi}{2}$.
Further, scaling by $\frac{2}{\pi}$ fits the ball into $[-1,1]^3$.
Encoding the sign of the last component as the chart index gives the compression function:
\begin{equation}
    \vect{f}_\mathrm{Exp}^{-1}(\uvec{x})=
    \begin{bmatrix}
        \vect{u}\\
        c
    \end{bmatrix}
    =
    \begin{bmatrix}
       \tau\,\uvec{x}_{1:3} \\
       \mathrm{neg}(\ucomp{x}_4)
    \end{bmatrix},\quad
    \tau=\frac{\frac{2}{\pi}\arctan{\frac{\sqrt{1-\ucomp{x}_4^2}}{\vert \ucomp{x}_4 \vert}}}
              {\sqrt{1-\ucomp{x}_{4}^2}},
\end{equation}
where $\tau=\frac{2}{\pi}$ for $\ucomp{x}_4=\pm1$.
The decompression function is
\begin{equation}
\vect{f}_\mathrm{Exp}(\vect{u};c)=
\uvec{x}=
\begin{bmatrix}
    \sin\left(\upsilon\right)\frac{\vect{u}}{\Vert\vect{u}\Vert_2}\\
    (-1)^c\cos\left(\upsilon\right)
\end{bmatrix},
\quad
    \upsilon=\frac{\pi}{2}\Vert \vect{u}\Vert_2.
\end{equation}

\subsection{Error Analysis}\label{Sec:ErrorAnalysis}
\begin{table}
    \begin{center}
    \footnotesize
    \input{tabs/ParameterizationMethodsSummary.tex}
    \end{center}
    \caption{\emph{Parameterization Method Summary.}
    For each method (first row), we list the number of \emph{charts}.
    Superscripts distinguish results for domain and range quantization.
    The rows $\Delta Q$ show the first-order Taylor approximation of the maximum geodesic error.
    The rows $\vect{u}$ show one domain point where the respective maximum geodesic error is reached.
    The rows $\kappa$ show the compactness factor.
    }\label{tab:ParameterizationMethodsFacts}
\end{table}
We find the \emph{maximum} quantization error of \cref{Eq:QuantizationError} for each parameterization.
For the proofs, see the \SupplementalName.
We choose the maximum over the mean or median, because it is independent of any particular input distribution.
\Cref{tab:ParameterizationMethodsFacts} summarizes the maximum first-order errors $\Delta Q^{\mathrm{domain}}$ and $\Delta Q^{\mathrm{range}}$ and their locations.
We verify the results numerically in \cref{sec:results-codecs}.

As with 3D unit vectors~\cite{Meyer12RTG}, $\lTwo$ is ill-conditioned for quaternions.
The derivative of $x_4=\sqrt{1-\Vert\vect{u}\Vert_2^2}$ is unbounded at the equator, which makes the maximum error $\mathcal{O}\left({\sqrt{\varepsilon}}\right)$.
Its maximum error is not only worse than that of all other methods, but even worse than quantizing all \emph{four} components of a quaternion directly at the same bit budget.
See \cref{tab:tangent-frame-quality} for measurements.
In floating point, the cancellation in $1-\Vert\vect{u}\Vert_2^2$ adds a further error.
Therefore, we omit $\lTwo$ from the following discussion.

To compare methods at equal bit budgets, we compute a \emph{compactness factor} $\kappa$, which accounts for the number of charts $C$:
packing a unit quaternion into $k$ bits leaves $U\approx(2^k/C)^{1/3}$ samples per parameter, hence $\Delta Q\approx\kappa\cdot 2^{-k/3}$, where $\kappa$ is $2\,C^{1/3}$ times the factor before $\varepsilon$, converted to degrees.
This makes $\kappa$ human-readable: dividing it by $2^{k/3}$ gives the maximum error in degrees, 
e.g., $\qty{156}{\degree}\cdot 2^{-32/3}\approx\qty{0.096}{\degree}$ for $\lInfTan$ at \num{32} bits.
$\lInfTan$ consistently outperforms all other methods.
One reason is that the warp function dampens the worst-case error around the poles.
Although this increases errors elsewhere, they remain below this new, lower maximum.
Another reason is \emph{domain utilization}:
All ${\vect{u}}\in\left[-1,1\right]^3$ of $\lInfTan$ and $\lInf$ map to unit vectors.
For $\lOne$ and $\lTwoI$, this is not the case, which wastes memory.
$\lTwoI$ comes close to $\lInfTan$ when we use the more complex range quantization.

It is surprising that $\lOne$ performs the worst among the analyzed methods.
This contrasts with the study of 3D unit vectors~\cite{Meyer10OFN}:
There, an $\lOne$-parameterization variant, octahedral normal vectors, performs best.
Octahedral normal vectors obtain perfect domain utilization by unwrapping the octahedron to a plane.
For unit quaternions, domain utilization of $\lOne$ is poor, and we did not find a transform from the octahedron-shaped domain to an optimal cube-shaped domain.
Instead, we subdivide the domain into 16 tetrahedra.
We enumerate only valid samples inside each tetrahedron with a unique integer index by adapting \emph{optimal simplex sampling}~\cite{Kuth21VBA}.
Then a quaternion is represented as a four-bit tetrahedron index and a $(k-4)$-bit index inside each tetrahedron, where $k$ is the number of bits for a compressed quaternion.
Since a tetrahedron occupies $1/6$ of its enclosing cube, $2^{k-4}$ states per tetrahedron correspond to an equivalent cube quantized with $6\cdot 2^{k-4}$ states.
This results in a sample spacing of approximately $(6\cdot 2^{k-4})^{-1/3}$ when using optimal simplex sampling~\cite{Kuth21VBA}.
It reduces the quantization error of $\lOne$ by a factor of $6^{-1/3}$, i.e., to $\kappa^\mathrm{domain}\approx\qty{275}{\degree}$ and $\kappa^\mathrm{range}\approx\qty{159}{\degree}$ (cf.\ \cref{tab:ParameterizationMethodsFacts}).
This is still worse than $\lInfTan$ with $\kappa\approx\qty{156}{\degree}$.
Furthermore, its derivation and implementation are tedious and involve numerical pitfalls and cube roots~\cite{Kuth21VBA}.
This contrasts with the simplicity of the other methods.

%% file: codes/PackUnpack.tex
\begin{lstlisting}[language=C++, basicstyle=\small\ttfamily, morekeywords={uint}, mathescape=true]
uint pack(uint $u$[], uint $c_1$, uint $c_2$) {
    uint code = $c_1$ * $C_2$ + $c_2$;
    for (int i = 2; i >= 0; --i)
        code = code * $U$ + $u$[i];
    return code;
}
void unpack(uint code, uint $u$[],
            uint& $c_1$, uint& $c_2$) {
    for (int i = 0; i < 3; ++i, code /= $U$)
        $u$[i] = code % $U$;
    $c_2$ = code % $C_2$; $c_1$ = code / $C_2$;
}
\end{lstlisting}

%% file: tabs/ParameterizationMethodsSummary.tex
\begin{tblr}{
  colspec = {l X[c] X[c] X[c] X[c] X[c]},
  colsep = 1pt,
  width = \linewidth,
}
    \toprule
    & $\lOne$ & $\lTwoI$ & $\lInf$ & $\lInfTan$ & $\mathrm{Exp}$ \\
    \midrule
    \#Charts          & $2$ & $8$ & $8$ & $8$ & $2$ \\
    \midrule
    $\vect{u}^{\mathrm{domain}}$  & $\frac{1}{4}[1,1,1]^\top$ & $\frac{1}{2}[1,1,1]^\top$ & $[0,0,0]^\top$ & $[0,0,0]^\top$ & $[0,0,0]^\top$ \\
    $\Delta Q^{\mathrm{domain}}$ [\si{\radian}] & $2\sqrt{3}\,\varepsilon$ & $\frac{\sqrt{6}}{2}\varepsilon$ & $\frac{\sqrt{3}}{2}\varepsilon$ & $\frac{\pi\sqrt{3}}{8}\varepsilon$ & $\frac{\pi\sqrt{3}}{4}\varepsilon$ \\
    $\kappa^{\mathrm{domain}}$ [\si{\degree}] $\downarrow$ & $500$ & $281$ & $198$ & \textbf{156} & $196$ \\
    \midrule
    $\vect{u}^{\mathrm{range}}$   & $\frac{1}{4}[1,1,1]^\top$ & $[\frac{\sqrt{2}}{2},0,0]^\top$ & $[0,0,0]^\top$ & $[0,0,0]^\top$ & $[0,0,0]^\top$ \\
    $\Delta Q^{\mathrm{range}}$ [\si{\radian}] & $2\,\varepsilon$ & $\frac{\sqrt{2}}{2}\varepsilon$ & $\frac{\sqrt{3}}{2}\varepsilon$ & $\frac{\pi\sqrt{3}}{8}\varepsilon$ & $\frac{\pi\sqrt{3}}{4}\varepsilon$ \\
    $\kappa^{\mathrm{range}}$ [\si{\degree}] $\downarrow$ & $289$ & $162$ & $198$ & \textbf{156} & $196$ \\
    \bottomrule
\end{tblr}

%% file: secs/B4warpfunctions.tex
\section{Warp Functions}\label{Sec:WarpFunctions}
% Figure from doc/WarpFunctionFigure/WarpFunctionFigure.ipynb with FIGURE = "measured_bits" and Y_AXIS = "kappa".
\begin{figure}[t]
    \centering
    \includesvg[inkscapearea=page,width=\columnwidth]{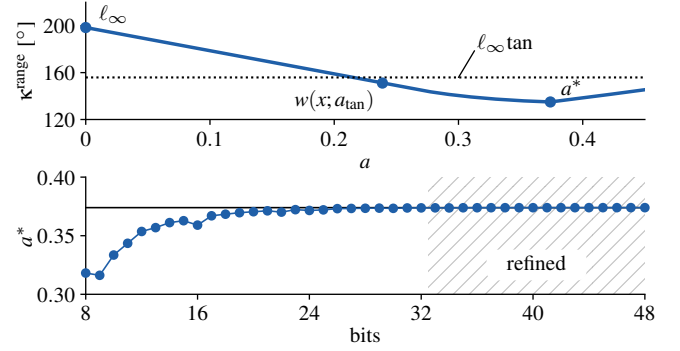}
    \caption{\emph{Compactness Factor $\kappa^\mathrm{range}$ and Optimum $a^\ast$} for unit quaternions.
    Top: $\kappa^\mathrm{range}$ of the cubic warp $w(x;a)$ from the maximum range quantization error over all cells, measured on the codec with \num{32} bits.
    The dotted line marks $\lInfTan$, which does not depend on $a$.
    Bottom: the optimum $a^\ast$ of each bit budget, by brute force up to \num{32} bits and refined from the previous budget above (hatched).
    The bit budgets omit the sign $c_2$, so \num{32} bits give $U=\num{1024}$, whereas $\kappa^\mathrm{range}$ counts it, as in \cref{tab:ParameterizationMethodsFacts}.
    }\label{fig:WarpKappa}
\end{figure}
Warping the domain interval as in $\lInfTan$ yields the lowest error; however, it requires an expensive tangent evaluation.
Therefore, we approximate the tangent with a cubic polynomial:
\begin{equation}
    w(x;a)=a\cdot x^3 + \left(1-a\right)\cdot x.\label{Eq:CubicWarpFunction}
\end{equation}
To keep neighboring charts seamless, we impose $w(-1)=-1$, $w(0)=0$, and $w(1)=1$, leaving a single parameter $a$.
One option to determine $a$ is to approximate the tangent with a Remez algorithm.
The Remez algorithm finds the polynomial of minimal maximum deviation from a function over an interval, and we get $a_\mathrm{tan}=\num{0.238798}$.
However, the tangent is only optimal for 2D unit vectors; thus, $a_\mathrm{tan}$ is not necessarily a good choice for quaternions.
Hence, we must find $a$ that minimizes $\kappa$.

\subsection{Finding the Optimum on the Grid}
First, we choose a target bit size $k$ for our packed quaternion.
For that bit size, we determine $\kappa(a)$ using a brute-force approach:
We iterate over all cells of the quantization domain grid, find the maximum geodesic error of each cell (see \cref{Sec:FindTheOptimumInACell}), and finally take the maximum over all cells.
By the symmetry of the chart in the signs and the order of its axes, it suffices to score the cells in $[0,1]^3$ with $u_1\le u_2\le u_3$, i.e.\ $1/48$ of the chart.

We start with 1000 samples for $a\in[0, 1)$ and take the best $a'$.
We take a subinterval around $a'$ whose interval boundaries are neighboring samples of $a'$.
Then we repeat this process with the new subinterval until the interval becomes smaller than \num{e-8}.
This gives a numerically optimal $a^\ast(k)$ for a given bit size.
From that we compute $\kappa(a)$.
For 32 bits and range quantization, \cref{fig:WarpKappa} top shows how $\kappa^{\mathrm{range}}(a)$ changes over $a$.

This search takes roughly eight minutes on a GPU, so it is feasible for 8 to 32 bits.
For larger bit sizes, we exploit that $a^\ast(k)$ appears to converge to a fixed value as shown in \cref{fig:WarpKappa} bottom.
To keep computation times manageable, we use a refinement approach: we sample $a$ at 11 points in the interval $[a^\ast(k-1)-0.01, a^\ast(k-1)+0.01]$, rather than in $[0,1)$.
In addition, instead of checking all cells, we iteratively go from a coarse grid to a finer grid.

\subsection{Find the Maximum Geodesic Error of a Cell}\label{Sec:FindTheOptimumInACell}
The maximum geodesic error inside a cell lies at the circumcenter of four cell corners forming a Delaunay tetrahedron.
This circumcenter on the hypersphere is given by the null-space of the differences of the four decoded corners, i.e., by their 4D cross product, normalized to unit length.
Out of the $\tbinom{8}{4}=70$ possible tetrahedra, we only consider those non-degenerate, with an empty circumsphere, and with a circumcenter inside the cell domain.
The largest circumradius is the maximum geodesic error of the cell.
For an arbitrary point set, the maximum could also lie on a face or an edge of the cell, but our experiments show that adding these candidates never changes the result.

Cells at chart boundaries, i.e., for $u_i = 1$, need special treatment: the cell is only half as wide and has no corners beyond the boundary.
Its maximum can hence lie on the boundary face, so we pin the respective axis to $1$ and let this constraint take the place of one equidistance constraint of the circumcenter.

\subsection{Results}
The optima converge to $a^\ast=\num{0.373964}$, which minimizes $\kappa^\mathrm{range}$.
Although each $k$ has an optimal $a^\ast$, we use one $a^\ast$ for all bit budgets, as it increases the maximum error by at most \qty{1.4}{\percent} from \num{16} bits on and \qty{0.23}{\percent} from \num{24} bits on, and call the resulting parameterization $\lInfCub$.
At 32 bits, this results in $\Delta Q^\mathrm{range}=0.5889\,\varepsilon$ and $\kappa^\mathrm{range}=\qty{135}{\degree}$, \qty{10.7}{\percent} below $w(x;a_\mathrm{tan})$ and \qty{13.4}{\percent} below $\lInfTan$ (cf.\ \cref{tab:ParameterizationMethodsFacts}).
% Evaluating every cell, we verify that the maximum error approaches this limit from above at every bit budget up to \num{32} bits, decreasing from $0.6112\,\varepsilon$ at \num{8} bits over $0.5947\,\varepsilon$ at \num{16} bits to $0.5889\,\varepsilon$ at \num{32} bits.
For domain quantization, $a^{\ast\mathrm{domain}}=\num{0.241752}$ is optimal, with $\kappa^\mathrm{domain}=\qty{152}{\degree}$.

In the \SupplementalName, we show that any code with $2^k$ states has a maximum error of at least $\kappa^\mathrm{lower}\cdot 2^{-k/3}$ with $\kappa^\mathrm{lower}\approx\qty{96}{\degree}$.
Hence, for large bit budgets, $\lInfCub$ needs only about $3\log_2\frac{\kappa^{\lInfCub,\mathrm{range}}}{\kappa^\mathrm{lower}}\approx\num{1.5}$ bits more than this lower bound.

% As for the methods in \cref{tab:ParameterizationMethodsFacts}, the maximum error is proportional to the sample spacing, but for the cubic warp, we have not derived a closed form of its compactness factor $\kappa^\mathrm{range}(a)$.

%% file: secs/B5results.tex
\section{Results and Discussion}\label{Sec:Evaluation}
In \cref{sec:results-codecs}, we empirically confirm our error analysis from \cref{Sec:ErrorAnalysis} and analyze the computational complexity of decoding the different parameterizations.
Next, in \crefrange{sec:results-animation}{sec:results-gaussians}, we evaluate our $\lInfCub$ in skeletal animation, tangent frames, and Gaussian splatting.
For each, we compare against uncompressed quaternions and the original codec in use, report angular and image errors, and, for animation and Gaussian splatting, the runtime cost.
Gaussian splatting results were measured on an \nvidiaGpu{}, the rest on an \amdGpu{} GPU.
A reference C implementation of $\lInfCub$ is available at
\ifthenelse{\equal{\BlindSubmission}{1}}
{
    \url{https://Removed.Due/To/Blind/Submission}
}
{
\url{https://github.com/GeoFlowResearch/UnitQuaternionCompressionSupplemental}
}.

\subsection{Quantization Error and Decode Cost}\label{sec:results-codecs}
For \cref{tab:decode-cycles}, we encode and decode $2^{24}$ random quaternions from Shoemake's algorithm~\cite{Shoemake92URR} with all analyzed parameterizations at various bit budgets, under both domain and range quantization schemes, and compare the measured angular error against the predictions of \cref{Sec:ErrorAnalysis}.
All values align with the analytic predictions.
Across all budgets and both quantization schemes, $\lInfCub$ attains the lowest maximum error of all methods, at the price of a mean error at most \qty{2}{\percent} above $\lInfTan$.

To measure GPU decode compute cost, we use our C implementation in AMD's HIP.
For a fair comparison, we implement every other parameterization in the same style and apply the same kinds of optimizations to each decoder.
We measure decode cost in GPU cycles by reading the hardware cycle counter before and after a single decode.
We prevent the compiler from reordering or eliminating the code in between, and subtract the cost of an empty function.

For reference, a single slerp between two quaternions costs \num{171} cycles on the same GPU, so at $\le$\num{32} bits every decoder except $\lInfTan$ is cheaper than one interpolation.
Unpacking is cheaper whenever the payload splits evenly into three and radix packing needs no integer division, which is the case at \num{16} bits for $\mathrm{Exp}$ using a single chart bit, and at \num{24} and \num{48} bits for the methods with three chart bits.
$\mathrm{Exp}$ and $\lInfTan$ decoders evaluate trigonometric functions and tend to be the most expensive, while all purely algebraic decoders stay at or below \num{135} cycles.
Thus, $\lInfTan$, whose warp is compiled to $\tan x = \frac{\sin x}{\cos x}$, costs \numrange{80}{159} cycles more than the unwarped $\lInf$ at every budget, whereas our cubic warp can be combined with the dequantization of \cref{Sec:Quantization} into three fused multiply-adds.

\begin{table*}[t]
\centering
\small
\input{tabs/decode_cycles.tex}
\caption{
    Angular error and decode cost of each parameterization at common bit budgets, measured over uniformly random unit quaternions.
    $\Delta Q$ is the asymptotic error prediction $\kappa\,2^{-k/3}$ of \cref{Sec:ErrorAnalysis}.
    Bold marks the best value per column within each budget.
}\label{tab:decode-cycles}
\end{table*}

%-----------------------------------------------------------------------------------------------------------------------
\subsection{Skeletal Animation}\label{sec:results-animation}

\newsavebox{\skinnedFramesPanels}
\newlength{\skinnedFramesHeight}
\begin{figure*}[t]
\centering
\sbox{\skinnedFramesPanels}{%
\begin{minipage}[b]{0.92\textwidth}
\centering
\subcaptionbox{\label{fig:skinned-frames-dragon-l2i}}{%
\includesvg[inkscapearea=page,width=\linewidth]{figs/skinned_frames_dragon_l2i48}}\\[2pt]
\subcaptionbox{\label{fig:skinned-frames-dragon-liw}}{%
\includesvg[inkscapearea=page,width=\linewidth]{figs/skinned_frames_dragon_liw40}}
\end{minipage}}%
\setlength{\skinnedFramesHeight}{\ht\skinnedFramesPanels}%
\usebox{\skinnedFramesPanels}\hfill
\includesvg[inkscapearea=page,height=\skinnedFramesHeight]{figs/skinned_frames_dragon_colorbar}
\caption{Skinning error on four keyframes of the flying animation of the demon dragon rendered at \numproduct{2048 x 2048}.
The reference uses 32-bit floats for each of the four quaternion components.
(a)~Ozz's $\lTwoI$-48 baseline with \num{48} bits per quaternion. 
(b)~Our $\lInfCub$-40 with \num{40} bits per quaternion. 
Colors show the per-pixel error $e=\max\left(|\Delta R|,|\Delta G|,|\Delta B|\right)$ to the reference image in 8-bit display levels, on the scale at the right; grey indicates $e=0$. 
Annotations report the percentage of covered pixels within one level, the percentage of exactly matching pixels, and PSNR. 
\emph{Demon Dragon} by \href{https://sketchfab.com/endlessvoidmc}{EndlessVoidMC}, \href{https://creativecommons.org/licenses/by/4.0/}{CC~BY~4.0}.}
\label{fig:skinned-frames-dragon}
\end{figure*}

We integrated $\lInfCub$ into the open-source animation library \emph{ozz-animation}~\cite{ozzAnimation}, called Ozz in the following, replacing only its quaternion compression scheme.
Ozz uses a variant of $\lTwoI$ with 48 bits per quaternion:
\num{15} bits on each of the three stored components, two bits on the chart index, and one bit on the sign.

Since $\uvec{x}$ and $-\uvec{x}$ represent the same rotation, the sign bit can be omitted, and there are libraries that do so~\cite{Frechette17UE4}.
However, Ozz does not exploit this:
For Ozz's $\lTwoI$ variant without the sign bit, \num{46} bits remain, which three integer fields cannot share equally.
Therefore, Ozz keeps the sign bit and uses \num{15} bits per component.
In contrast, our mixed-radix packing of \cref{Sec:PackAndUnpack} allows fractional numbers of bits per component, so we can omit storing the sign bit.
Instead of $2^{15}=\num{32768}$ states per component, we use $\lfloor 2^{46/3}\rfloor=\num{41285}$.

We evaluate our two sign-free configurations: $\lInfCub$-48 as a storage-equivalent replacement, and $\lInfCub$-40 to demonstrate the effect of saving one byte per quaternion.
Sign-storing variants of both isolate the effect of the recovered bit.
We measure the geodesic error between decoded and uncompressed quaternions over all keyframes of the Mixamo corpus~\cite{MixamoDataset}.

\begin{table}
\centering
\small
\input{tabs/mixamo_quality_sampling.tex}
\caption{Results on all \num{34884287} keyframes of the Mixamo corpus.
We compare various \emph{codec}s using different numbers of \emph{bits}.
\emph{States} shows the number of states per component and \emph{sign bit} indicates whether the sign bit is stored.
The column \emph{max error} reports the maximum geodesic error.
The column \emph{sampling} shows the time for \num{1024} skinned characters, per frame as the mean over \num{60} frames of the fastest of \num{4} runs.
Bold marks the best result.
}
\label{tab:mixamo-quality}
\end{table}

\Cref{tab:mixamo-quality} shows that at the same storage budget, $\lInfCub$-48 reduces the maximum error of Ozz's codec by a factor of \num{5.6}.
With the same \num{15} bits per component as Ozz, our parameterization and range quantization alone reduce the maximum error by a factor of \num{4.5}.
Omitting the sign bit gives $2^{1/3}$ more states per component and reduces it by another factor of \num{1.26}.
Ozz also uses domain quantization and an approximate square root, both of which increase its error.
Therefore, Ozz's error even exceeds our domain-quantization prediction for $\lTwoI$ in \cref{Sec:ErrorAnalysis}.

With the gained flexibility of mixed-radix packing, we can use almost any bit size to pack quaternions.
$\lInfCub$-40 trades accuracy for storage and saves \qty{33.3}{\mebi\byte} over the corpus.
Its maximum error is \qty{12}{\percent} above that of Ozz's $\lTwoI$-48, as shown in \cref{tab:mixamo-quality}.
However, in the rendering of \cref{fig:skinned-frames-dragon}, both configurations produce comparable image errors.

\Cref{tab:mixamo-quality} reports the sampling time for \num{1024} skinned characters.
$\lInfCub$ samples \qtyrange{12}{16}{\percent} faster than Ozz's $\lTwoI$-48 during playback.
Omitting the sign bit costs slightly more sampling time but improves accuracy.
Since sampling accounts for at most \qty{14.5}{\percent} of the CPU frame time, the codec is effectively free at runtime, and the benefit of $\lInfCub$ lies in accuracy and storage.

%-----------------------------------------------------------------------------------------------------------------------

\subsection{Tangent Frames}\label{sec:results-tangent-frames}
\begin{figure*}[tp]
\centering
\includesvg[inkscapearea=page,width=\textwidth]{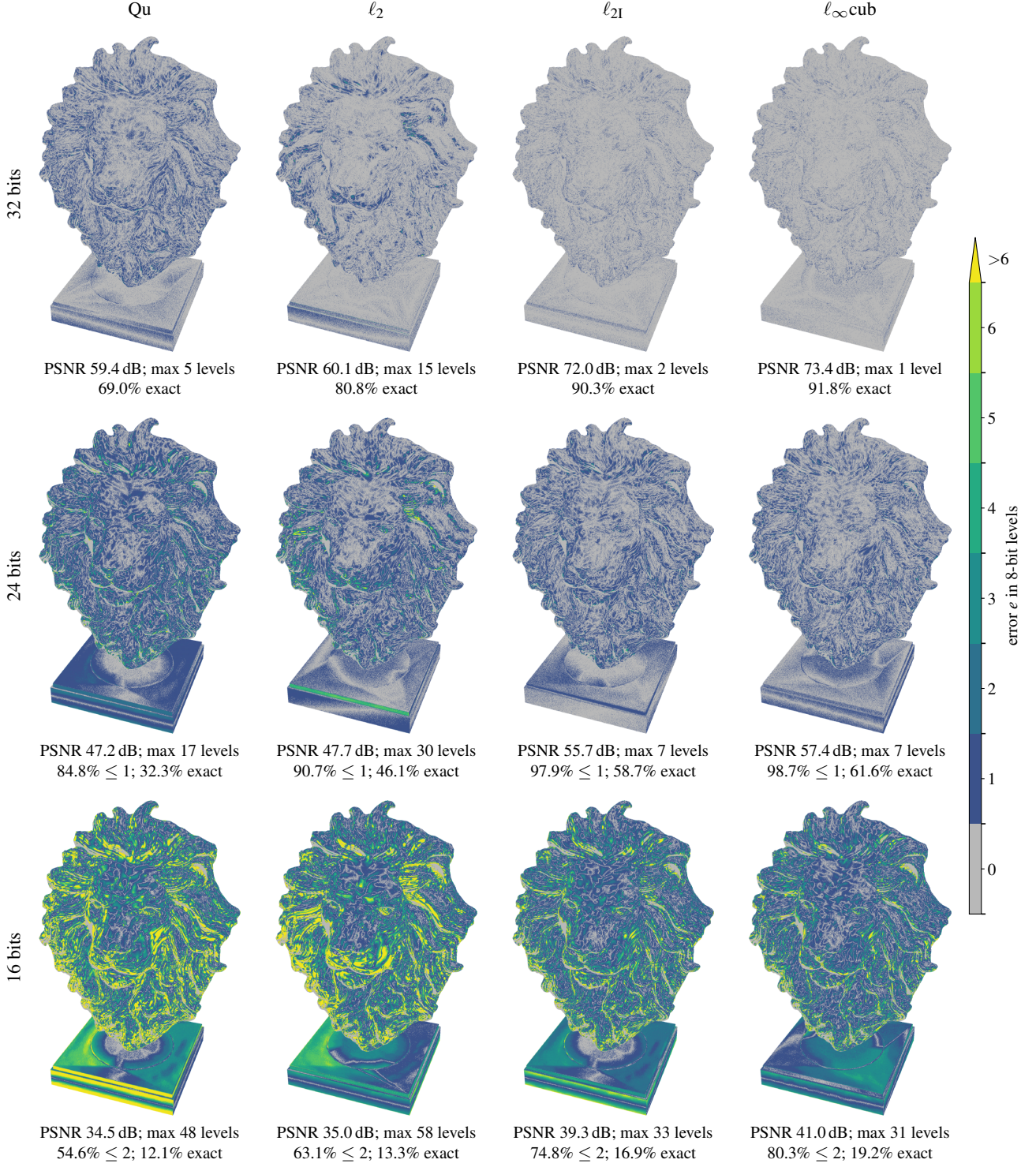}
\caption{Tangent frame compression error on the lion head model of Poly Haven (\numproduct{2048 x 2048} renders). 
Rows correspond to the bit budget and columns to the parameterization; $\mathrm{Qu}$ quantizes all four components. 
Colors show the per-pixel error $e=\max\left(|\Delta R|,|\Delta G|,|\Delta B|\right)$ to the reference render with uncompressed tangent frames, shown at the top right of \cref{fig:teaser}, in 8-bit display levels on the scale at the right; grey indicates $e=0$, and errors above \num{6} 
Annotations report the percentage of covered pixels with $e\le1$ at \num{24} bits and $e\le2$ at \num{16} bits, the percentage of exactly matching pixels, the maximum error, and PSNR. 
Since the error of a single view varies by almost three orders of magnitude with the view direction, \cref{tab:tangent-frame-quality} reports metrics over \num{64} views per asset.}
\label{fig:tangent-frames}
\end{figure*}

Normal mapping requires a tangent frame at every shaded fragment, i.e., a rotation, possibly combined with a reflection.
Since $\uvec{x}$ and $-\uvec{x}$ represent the same rotation, we can save one chart bit, which we reuse to store the handedness.
Tangent frames are stored per vertex, read in every draw call, and decoded independently.
We implemented all parameterizations in a Direct3D~12 renderer that encodes the tangent frames on the CPU and decodes them in the vertex shader.

We compare range quantization for $\lTwo$, $\lTwoI$, and $\lInfCub$ at \num{32}, \num{24}, and \num{16} bits per tangent frame.
$\mathrm{Qu}$ naively quantizes all four components with domain quantization.
With our quantization, we can never decompress $\ucomp{x}_4=0$ for an even number of states $U$, so we can encode the reflection in the sign of $\ucomp{x}_4$~\cite{Frey11SSD}.
We evaluate on \num{14} glTF assets with \num{397328} tangent frames from the Khronos glTF Sample Assets, Poly Haven, and Smithsonian Open Access.
\num{13} of the \num{14} assets lack tangents, so we derive them from the texture coordinates.
Besides the geodesic error, we compare renders with compressed and uncompressed tangent frames by PSNR and LDR-FLIP~\cite{Andersson20FLIP} over \num{64} views per asset, distributed on a Fibonacci sphere.
All renders use a constant grey albedo, so that textures cannot mask errors.
Normals are sampled from the asset's tangent-space normal map.

\begin{table}
\centering
\small
\setlength{\tabcolsep}{5pt}
\input{tabs/tangent_frame_quality.tex}
\caption{Compressing \num{397328} tangent frames from \num{14} glTF assets. 
For various \emph{method}s, we use different bit \emph{budget}s.
\emph{Bits used} is the base-2 logarithm of the number of codes, which falls short of the budget where that number is not a power of two. 
The \emph{quaternion} columns show the \emph{max} and \emph{mean} geodesic error over all frames against the uncompressed frames.
\emph{PSNR} and \emph{FLIP} are pooled over \num{64} views per asset at \numproduct{1920 x 1080}.
FLIP is evaluated at \num{67} pixels per degree. 
Bold marks the best method per budget.}
\label{tab:tangent-frame-quality}
\end{table}

\Cref{tab:tangent-frame-quality} shows that at every budget, $\lInfCub$ is the most accurate parameterization in all four metrics.
Compared with $\lTwoI$, it reduces the maximum error by a factor of \numrange{1.18}{1.23}, which translates into \qtyrange{1.1}{1.3}{\decibel} higher PSNR at the same storage.
$\mathrm{Qu}$ spends about a quarter of its bits on a redundant fourth component, so its errors are \numrange{2}{5} times those of $\lInfCub$ throughout.
$\lTwo$ has a moderate mean but a large maximum error.
Even $\mathrm{Qu}$ has a lower maximum error than $\lTwo$.

At \num{32} bits, the rendered error of $\lInfCub$ stays within one or two display levels over most of the model (\cref{fig:tangent-frames}), whereas at \num{16} bits, only about a fifth of the pixels match the uncompressed render exactly, although \qty{80}{\percent} stay within two levels.
We use nine 32-bit floats to represent uncompressed tangent frames.
On our GPU, tangent-frame decompression did not affect the frame time in a measurable way.

\subsection{Gaussian Splatting}\label{sec:results-gaussians}
\begin{figure*}[tp]
\centering
\includesvg[inkscapearea=page,width=\textwidth]{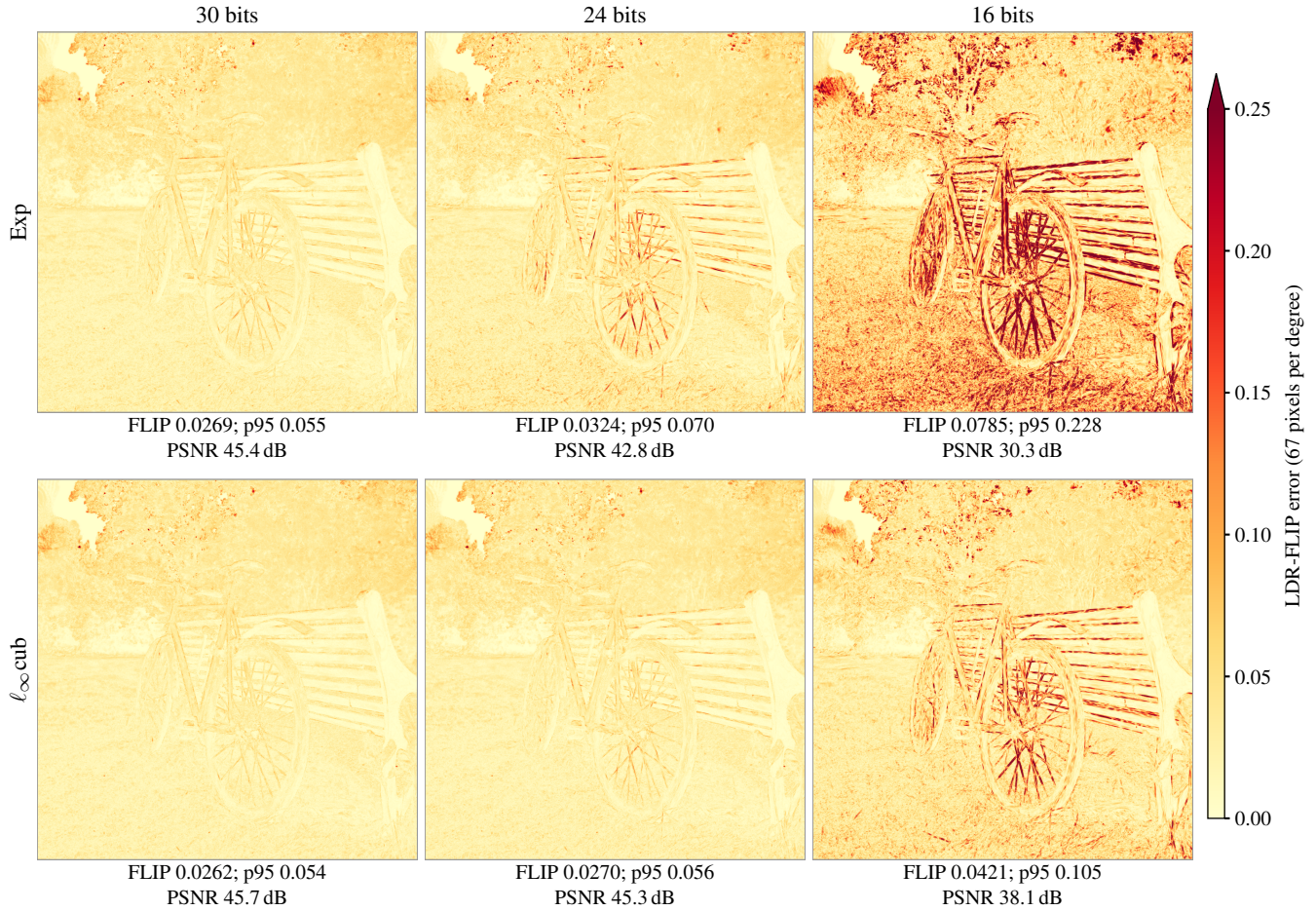}
\caption{Gaussian splatting compression error on the \emph{bicycle} scene of Mip-NeRF~360~\cite{Barron22MipNeRF360} (\numproduct{2048 x 2048} renders, \num{1024} samples per pixel). Columns correspond to the bit budget and rows to the method. Colors show the per-pixel LDR-FLIP error to the reference render with uncompressed rotations, shown at the bottom of \cref{fig:teaser}, on the scale at the right; white indicates zero error. Annotations report the mean and 95th percentile (p95) of LDR-FLIP, and PSNR. \Cref{tab:gs-rotation-quality} reports metrics over all three scenes. Reconstruction by \href{https://superspl.at/user/seeget3d}{Seeget3D}, \href{https://creativecommons.org/licenses/by/4.0/}{CC~BY~4.0}.}
\label{fig:gaussian-splatting-bike}
\end{figure*}

3D Gaussian splatting~\cite{Kerbl233DG} represents a scene by millions of anisotropic Gaussians, each oriented by a unit quaternion.
Compression methods for such scenes~\cite{Bagdasarian25ZIP} either cluster the rotations together with other attributes into code books~\cite{Niedermayr24C3G,Lee24C3G} or quantize each rotation individually~\cite{Xie24MGS,NianticSPZ,PlayCanvasSOG,rijsdijk2026GPS}.
Our method belongs to the second group, which keeps every Gaussian independently decodable.

We integrate our method into the CUDA renderer of Gaussian point splatting~\cite{rijsdijk2026GPS}, which packs the geometry of each Gaussian into \num{128} bits with \num{30} bits for the quaternion.
They use exponential mapping~\cite{Reynolds17QQ} for quaternion compression (Exp).
Note that their Exp implementation uses incorrect rounding when quantizing, unlike our reference implementation, producing slightly worse results.
We replace Exp with our $\lInfCub$ method at \num{30}, \num{24}, and \num{16} bits.
Analogously to animation keys, we can drop the sign, since $\uvec{x}$ and $-\uvec{x}$ yield the same Gaussian.
% $\lTwoI$ as used by SPZ~\cite{NianticSPZ}, which is proposed as the compression format of the Khronos glTF Gaussian splatting extension~\cite{KhronosGSSPZ},
We evaluate on the \emph{bicycle}, \emph{garden}, and \emph{room} scenes of Mip-NeRF~360~\cite{Barron22MipNeRF360}, each rendered from five close and distant viewpoints at \numproduct{2048 x 2048} pixels with \num{1024} samples per pixel.

\begin{table}[t]
\centering
\small
\setlength{\tabcolsep}{5pt}
\input{tabs/gaussian_splats_quality_reduced.tex}
\caption{Gaussian rotation compression on the \emph{bicycle}, \emph{garden}, and \emph{room} scenes. 
We compare two \emph{method}s at different bit \emph{budget}s.
The column \emph{quaternion} shows the geodesic errors against the uncompressed quaternion, with the maximum and mean over all Gaussians. 
\emph{PSNR} and \emph{FLIP} are averaged over five views per scene. Bold marks the better method per budget.}
\label{tab:gs-rotation-quality}
\end{table}

\Cref{tab:gs-rotation-quality} shows that at the same budget $\lInfCub$ reduces the angular errors and improves image quality compared to Exp.
The reference implementation with Exp achieves \qty{45.9}{\decibel} at \num{30} bits, while $\lInfCub$ nearly achieves this quality at \num{24} bits.
This lead increases at \num{16} bits with an advantage of \qty{6.67}{\decibel}.
This also holds for \FLIP{}, as shown in \cref{fig:gaussian-splatting-bike}.
At equal budgets, we did not observe any measurable difference in render time; decoding the rotations takes about \qty{0.5}{\milli\second} per frame for all of them, compared with \qty{11.8}{\milli\second} for rendering.

%% file: tabs/decode_cycles.tex
\begin{tabular*}{\textwidth}{@{\extracolsep{\fill}}llrrrrrrr@{}}
\toprule
 & & & \multicolumn{6}{c}{Quaternion error [\si{\degree}]} \\
\cmidrule(lr){4-9}
 & & & \multicolumn{3}{c}{Domain quantization} & \multicolumn{3}{c}{Range quantization} \\
\cmidrule(lr){4-6}\cmidrule(lr){7-9}
\begin{tabular}[b]{@{}l@{}}Budget\\{[bits]}\end{tabular}
& Method & Cycles $\downarrow$ & Mean $\downarrow$ & Max $\downarrow$ & $\Delta Q$
& Mean $\downarrow$ & Max $\downarrow$ & $\Delta Q$ \\
\midrule
% gfx1100, full decoders, 524288 wave decodes, 16777216 quaternions each
% reference: 4x4 view-projection times a position, 25 cycles
% reference: 3x3 rotation times a tangent frame, 34 cycles
% reference: slerp between two quaternions, 171 cycles
\multirow{5}{*}{16}  % narrow 32-bit codes
 & $\lTwoI$           &            132 &           2.157171 &           7.429381 &   6.962015 &           2.118236 &           3.983630 &   4.019521 \\
 & $\mathrm{Exp}$     &            129 &           2.346221 &           4.820204 &   4.871393 &           2.319461 &           4.818251 &   4.871393 \\
 & $\lInf$            &            108 &           2.004001 &           4.864108 &   4.922888 &           1.979029 &           4.831496 &   4.922888 \\
 & $\lInfTan$         &            188 &  \textbf{1.903187} &           3.821377 &   3.866427 &  \textbf{1.875452} &           3.821377 &   3.866427 \\
 & $\lInfCub$ (Ours)  &            103 &           1.904901 &  \textbf{3.772493} &        --- &           1.909772 &  \textbf{3.391181} &        --- \\
\midrule
% gfx1100, full decoders, 524288 wave decodes, 16777216 quaternions each
% reference: 4x4 view-projection times a position, 25 cycles
% reference: 3x3 rotation times a tangent frame, 34 cycles
% reference: slerp between two quaternions, 171 cycles
\multirow{5}{*}{24}  % narrow 32-bit codes
 & $\lTwoI$           &            103 &           0.337169 &           0.996873 &   1.096449 &           0.331288 &           0.606464 &   0.633035 \\
 & $\mathrm{Exp}$     &            157 &           0.369955 &           0.761963 &   0.767196 &           0.365739 &           0.755470 &   0.767196 \\
 & $\lInf$            &             80 &           0.313168 &           0.754806 &   0.775306 &           0.309210 &           0.754806 &   0.775306 \\
 & $\lInfTan$         &            239 &  \textbf{0.297412} &           0.598882 &   0.608924 &  \textbf{0.292833} &           0.597885 &   0.608924 \\
 & $\lInfCub$ (Ours)  &             67 &           0.297671 &  \textbf{0.589754} &        --- &           0.298241 &  \textbf{0.523886} &        --- \\
\midrule
% gfx1100, full decoders, 524288 wave decodes, 16777216 quaternions each
% reference: 4x4 view-projection times a position, 25 cycles
% reference: 3x3 rotation times a tangent frame, 34 cycles
% reference: slerp between two quaternions, 171 cycles
\multirow{5}{*}{32}  % narrow 32-bit codes
 & $\lTwoI$           &            135 &           0.053095 &           0.162046 &   0.172680 &           0.052169 &           0.096290 &   0.099697 \\
 & $\mathrm{Exp}$     &            155 &           0.058218 &           0.119754 &   0.120826 &           0.057554 &           0.119189 &   0.120826 \\
 & $\lInf$            &            115 &           0.049321 &           0.119526 &   0.122103 &           0.048694 &           0.119526 &   0.122103 \\
 & $\lInfTan$         &            273 &  \textbf{0.046844} &           0.094169 &   0.095900 &  \textbf{0.046116} &           0.093982 &   0.095900 \\
 & $\lInfCub$ (Ours)  &            111 &           0.046889 &  \textbf{0.092720} &        --- &           0.046965 &  \textbf{0.081655} &        --- \\
\midrule
% gfx1100, full decoders, 524288 wave decodes, 16777216 quaternions each
% reference: 4x4 view-projection times a position, 25 cycles
% reference: 3x3 rotation times a tangent frame, 34 cycles
% reference: slerp between two quaternions, 171 cycles
% \multirow{5}{*}{40}  % wide 64-bit codes
%  & $\lTwoI$           &            180 &           0.008361 &           0.025118 &   0.027195 &           0.008215 &           0.015270 &   0.015701 \\
%  & $\mathrm{Exp}$     &            139 &           0.009167 &           0.018894 &   0.019029 &           0.009063 &           0.018656 &   0.019029 \\
%  & $\lInf$            &            153 &           0.007766 &           0.018713 &   0.019230 &           0.007667 &           0.018713 &   0.019230 \\
%  & $\lInfTan$         &            256 &  \textbf{0.007378} &           0.014934 &   0.015103 &  \textbf{0.007263} &           0.014895 &   0.015103 \\
%  & $\lInfCub$ (Ours)  &            200 &           0.007384 &  \textbf{0.014651} &        --- &           0.007397 &  \textbf{0.012961} &        --- \\
% \midrule
% gfx1100, full decoders, 524288 wave decodes, 16777216 quaternions each
% reference: 4x4 view-projection times a position, 25 cycles
% reference: 3x3 rotation times a tangent frame, 34 cycles
% reference: slerp between two quaternions, 171 cycles
\multirow{5}{*}{48}  % wide 64-bit codes
 & $\lTwoI$           &             98 &           0.001317 &           0.003890 &   0.004283 &           0.001294 &           0.002384 &   0.002473 \\
 & $\mathrm{Exp}$     &            242 &           0.001444 &           0.002972 &   0.002997 &           0.001427 &           0.002934 &   0.002997 \\
 & $\lInf$            &             83 &           0.001223 &           0.002948 &   0.003029 &           0.001208 &           0.002948 &   0.003029 \\
 & $\lInfTan$         &            240 &  \textbf{0.001162} &           0.002355 &   0.002379 &  \textbf{0.001144} &           0.002339 &   0.002379 \\
 & $\lInfCub$ (Ours)  &             84 &           0.001163 &  \textbf{0.002295} &        --- &           0.001165 &  \textbf{0.002029} &        --- \\
\bottomrule
\end{tabular*}

%% file: tabs/mixamo_quality_sampling.tex
% Tabular only; the float and its caption are in secs/B5results.tex.
% Merges tabs/mixamo_quality.tex and tabs/skinned_frame_time.tex.
% Errors from ozz-animation-liw/paper/mixamo_compression/mixamo_compression.ipynb.
% Sampling from ozz-animation-liw build/temp/liw_skinned_benchmark_liwf_20261001
% (rendered, 1024 characters, 60Hz playback, best of 4 repetition means over 60 frames).
\begin{tabular}{@{}lrcrrr@{}}
\toprule
Method & Bits & \begin{tabular}[b]{@{}c@{}}Sign \\ bit\end{tabular} & States & \begin{tabular}[b]{@{}c@{}}Max error \\ {[\si{\degree}]} $\downarrow$\end{tabular} & \begin{tabular}[b]{@{}c@{}}Sampling \\ {[\si{\milli\second}]} $\downarrow$\end{tabular} \\
\midrule
Ozz's $\lTwoI$-48 & 48 & yes & \num{32768} & 0.00909 & 5.177 \\
$\lInfCub$-48 & 48 & yes & \num{32768} & 0.00203 & \textbf{\boldmath 4.371} \\
$\lInfCub$-40 & 40 & yes & \num{5160} & 0.01296 & 4.439 \\
\midrule
$\lInfCub$-48 & 48 & no & \num{41285} & \textbf{\boldmath 0.00161} & 4.538 \\
$\lInfCub$-40 & 40 & no & \num{6501} & 0.01020 & 4.492 \\
\bottomrule
\end{tabular}

%% file: tabs/tangent_frame_quality.tex
% Tabular only; the float and its caption are in secs/B5results.tex.
% Numbers from QuaternionCompression/doc/paper_tangent_frame/TangentFrameSection.ipynb (out/frame_quality.csv),
% every asset's object space half turned about y (OBJECT_ROTATION there).
% The code book row (16 bits) is left out; bold at 16 bits moves to $\lInfCub$.
\begin{tabular}{@{}llrrrrr@{}}
\toprule
 &  &  & \multicolumn{2}{c}{Quaternion [\si{\degree}]} & \multicolumn{2}{c}{Image} \\
\cmidrule(lr){4-5} \cmidrule(lr){6-7}
\begin{tabular}[b]{@{}l@{}}Budget\\{[bits]}\end{tabular} & Method & \begin{tabular}[b]{@{}r@{}}Bits\\used\end{tabular} & Max $\downarrow$ & Mean $\downarrow$ & PSNR $\uparrow$ & FLIP $\downarrow$ \\
\midrule
\multirow{4}{*}{32} & $\mathrm{Qu}$ & 32.00 & 0.4418 & 0.2210 & 58.47 & 0.00594 \\
 & $\lTwo$ & 32.00 & 1.5535 & 0.0830 & 63.91 & 0.00304 \\
 & $\lTwoI$ & 32.00 & 0.0996 & 0.0537 & 70.64 & 0.00216 \\
 & $\lInfCub$ & 32.00 & \textbf{\boldmath 0.0826} & \textbf{\boldmath 0.0475} & \textbf{\boldmath 71.71} & \textbf{\boldmath 0.00199} \\
\midrule
\multirow{4}{*}{24} & $\mathrm{Qu}$ & 24.00 & 1.7489 & 0.8943 & 46.36 & 0.01695 \\
 & $\lTwo$ & 24.00 & 4.0199 & 0.4524 & 50.27 & 0.01057 \\
 & $\lTwoI$ & 24.00 & 0.6180 & 0.3425 & 54.70 & 0.00831 \\
 & $\lInfCub$ & 24.00 & \textbf{\boldmath 0.5232} & \textbf{\boldmath 0.3019} & \textbf{\boldmath 55.76} & \textbf{\boldmath 0.00761} \\
\midrule
\multirow{4}{*}{16} & $\mathrm{Qu}$ & 16.00 & 7.4563 & 3.7285 & 34.13 & 0.04986 \\
 & $\lTwo$ & 16.00 & 10.5876 & 2.6412 & 36.23 & 0.03952 \\
 & $\lTwoI$ & 15.97 & 4.1652 & 2.2977 & 38.38 & 0.03447 \\
 & $\lInfCub$ & 15.97 & \textbf{\boldmath 3.3839} & \textbf{\boldmath 1.9354} & \textbf{\boldmath 39.71} & \textbf{\boldmath 0.03080} \\
\bottomrule
\end{tabular}

%% file: tabs/gaussian_splats_quality_reduced.tex
% Tabular only; the float and its caption are in secs/B5results.tex.
% Reduced to Exp and $\lInfCub$, as in the figure; the full table is tabs/gaussian_splats_quality.tex.
% Numbers from QuaternionCompression/doc/paper_gaussian_splatting/GaussianRotationSection.ipynb.
\begin{tabular}{@{}llrrrr@{}}
\toprule
 &  & \multicolumn{2}{c}{Quaternion [\si{\degree}]} & \multicolumn{2}{c}{Image} \\
\cmidrule(lr){3-4} \cmidrule(lr){5-6}
\begin{tabular}[b]{@{}l@{}}Budget\\{[bits]}\end{tabular} & Method & Max $\downarrow$ & Mean $\downarrow$ & PSNR [\si{\decibel}] $\uparrow$ & FLIP $\downarrow$ \\
\midrule
\multirow{2}{*}{30} & Exp & 0.3034 & 0.1533 & 45.90 & 0.02871 \\
 & $\lInfCub$ (ours) & \textbf{\boldmath 0.1059} & \textbf{\boldmath 0.0584} & \textbf{\boldmath 46.01} & \textbf{\boldmath 0.02851} \\
\midrule
\multirow{2}{*}{24} & Exp & 1.2016 & 0.6117 & 44.62 & 0.03141 \\
 & $\lInfCub$ (ours) & \textbf{\boldmath 0.4086} & \textbf{\boldmath 0.2334} & \textbf{\boldmath 45.83} & \textbf{\boldmath 0.02897} \\
\midrule
\multirow{2}{*}{16} & Exp & 7.7409 & 3.9188 & 35.01 & 0.06527 \\
 & $\lInfCub$ (ours) & \textbf{\boldmath 2.6355} & \textbf{\boldmath 1.4818} & \textbf{\boldmath 41.68} & \textbf{\boldmath 0.03822} \\
\bottomrule
\end{tabular}

%% file: secs/B6conclusion.tex
\section{Limitations}\label{Sec:Limitations}
Our analysis bounds the error over all unit quaternions and thus spends the same precision everywhere on $S^3$.
For cases where the rotations are concentrated, such as the keys of an animation track that covers only a small range of rotations, methods that adapt the quantization range per track or segment~\cite{frechette2020acl} can be more compact.
We derive closed-form first-order errors for every parameterization in \cref{tab:ParameterizationMethodsFacts}, including $\lInfTan$ with the tangent warp, but not for $\lInfCub$ with the cubic warp.
Range quantization makes encoding more complex than rounding, since it inverts the warp and evaluates the eight corners of a cell, but its cost remains $\mathcal{O}(1)$ per rotation.
Identity rotations, which are frequent in animations, are reconstructed exactly only when using an odd number of states $U$; this is trivial to enforce but not part of our uniform analysis.

\section{Conclusion and Future Work}\label{Sec:ConclusionAndFutureWork}
We analyzed the maximum error of common parameterizations for unit quaternions and advanced the state of the art by proposing a new method $\lInfCub$:
a combination of the $\ell_{\infty}$-norm, radix packing and range quantization, providing a lower maximum error.
Our evaluation confirmed this analysis empirically and showed that our decoding is computationally inexpensive.
Further, we showed that $\lInfCub$ can be used as a drop-in replacement in common graphics applications utilizing quaternions.

While this work focuses on unit quaternions ($n=4$), the parameterization is not specific to them.
For 3D unit vectors ($n=3$), the projection onto the cube, a cubic warp optimized for $S^2$, and mixed-radix packing yield a parameterization with six charts and two parameters.
In future work, we will compare it with octahedral encodings and use it for the normal of tangent frames stored as a normal and an angle (\cref{sec:PreviousWork}), packing the normal, angle, and handedness into a single mixed-radix code.
%We provide an analysis of the parameterizations commonly used to compress unit quaternions, including first-order approximations of their maximum quantization errors.
%Based on this analysis, we propose $\lInfCub$, which projects onto the unit hyper-cube and warps each parameter with a cubic polynomial whose single coefficient we %optimize numerically to minimize the maximum angular error.
%It has the lowest maximum error of all compared parameterizations and decodes with a few arithmetic instructions and without lookup tables.
%We combine $\lInfCub$ with range quantization and a mixed-radix packing into a complete codec.
%The packing stores the chart index and all quantized parameters as a single integer code, which, unlike separate bit fields, uses almost all states of a code word of %any size while giving every parameter the same precision.
%
%In all three of our evaluations (skeletal animation, normal mapping, and Gaussian splatting), our codec reduced the error of the existing rotation codec at the same %storage.
%It can thus serve as a drop-in replacement for existing rotation codecs.

%% file: secs/C0ack.tex
\section*{Acknowledgments}
\ifthenelse{\equal{\BlindSubmission}{0}}
{We thank Dominik Baumeister.}
{}
The animation clips of the ozz-animation evaluation are from Mixamo, as collected by Guo et al.~\cite{MixamoDataset}, and the \emph{Demon Dragon Full Texture} model is by EndlessVoidMC (CC~BY~4.0).
Of the tangent frame assets, \emph{DamagedHelmet} is by ctxwing after theblueturtle\_ (CC~BY~4.0), the rigging and animation of \emph{Fox} are by tomkranis and its glTF conversion by @AsoboStudio and @scurest (CC~BY~4.0), and \emph{BrainStem} is by Keith Hunter (Smith Micro Software), redistributed in the Khronos glTF Sample Assets under the Poser EULA.
The \emph{bicycle}, \emph{garden}, and \emph{room} scenes of Mip-NeRF~360~\cite{Barron22MipNeRF360} used in the Gaussian splatting evaluation were reconstructed by Seeget3D (CC~BY~4.0).

%% file: secs/S0proofs.tex
\section{AI Statement}
The following proofs were made with the help of the Computer Algebra Systems Maple 2026~\cite{Maple2026} and the AI assistant Claude (model Opus 5.5).
Prior work on 3D unit vectors~\cite{Meyer12RTG} served as a starting point.
The authors developed the proof strategies, verified every step by hand, reworked arguments into short proofs, verified the results numerically, wrote the text, and are responsible for the content.

\section{Errors for Quantized Quaternions}
When quantizing the components of an input quaternion $\uvec{x}$, each component is quantized on its own.
Using rounding, this changes each component by at most $\frac{\varepsilon}{2}$.
Therefore, the error vector $\vect{d}=\aprx{\uvec{x}}-\uvec{x}$ has length at most $\varepsilon$.
Since $\uvec{x}$ has unit length, the angle between $\uvec{x}$ and $\uvec{x}+\vect{d}$ is at most $\arcsin\Vert\vect{d}\Vert_2\leq\arcsin\varepsilon=\varepsilon+\mathcal{O}\left(\varepsilon^3\right)$.
Therefore, to first order
\begin{equation}
    \Delta Q^{\mathrm{Qu}}=\varepsilon.
\end{equation}
The bound is approach for $\uvec{x}=\frac{1}{2}[1,1,1,1]^\top$ if the grid lies such that the two components round up and two round down.

\section{A Lower Bound}\label{Sec:LowerBound}
An optimal unit vector representation distributes $N$ 4D unit vectors across $S^3$ such that their maximum geodesic error $\Delta Q^\mathrm{lower}$ is as small as possible.
We bound $\Delta Q^\mathrm{lower}$ from below following Alexa~\cite[Eq.~(17)]{Alexa22SFS}.
Every point of $S^3$ lies within $\Delta Q^\mathrm{lower}$ of one of the $N$ unit vectors.
Hence, the Voronoi cell of each unit vector lies inside the spherical cap of radius $\Delta Q^\mathrm{lower}$ around it, and the $N$ caps cover $S^3$.
Each cap has a volume of at most $\frac{4\pi}{3}\left(\Delta Q^\mathrm{lower}\right)^3$ and $S^3$ has a volume of $2\pi^2$, so $N\cdot\frac{4\pi}{3}\left(\Delta Q^\mathrm{lower}\right)^3\geq2\pi^2$ and
\begin{equation}
    \Delta Q^\mathrm{lower} \geq \left(\frac{3\pi}{2N}\right)^{\frac{1}{3}} 
    = \underbrace{\left(\frac{3\,\pi}{2}\right)^{\frac{1}{3}}}_{\kappa^\mathrm{lower}\approx \qty{96}{\degree}}\,2^{-\frac{k}{3}},
\end{equation}
where $N=2^k$ for a $k$-bit code, so $\kappa^\mathrm{lower}$ is directly comparable to the compactness factors in \cref{tab:ParameterizationMethodsFacts}.
No parameterization with $N$ codes can have a smaller maximum error.

\section{Error Analysis for Parameterization Methods}\label{Sec:ErrorAnalysisProof}
For each parameterization $\vect{f}$, we determine the locations $\vect{u}^\mathrm{domain}$ and $\vect{u}^\mathrm{range}$ where the domain error $\Delta Q^\mathrm{domain}$ and the range error $\Delta Q^\mathrm{range}$ are the largest.
Due to symmetry, we consider only non-negative $u_i$ on a single chart.
Thus, we omit the chart index.
Since $\vect{f}\in\mathbb{R}^{4}$ is typically computed from an unnormalized function $\vect{v}\left(\vect{u}\right)$, with $\vect{u}\in\mathbb{R}^{3}$ and $\vect{v}\in\mathbb{R}^{4}$, we write
\begin{equation}
    \label{Eq:DefineVSigma}
    \vect{f}(\vect{u})  = \mathrm{norm}_2\left(\vect{v}\left(\vect{u}\right)\right)
                        = {\vect{v}\left(\vect{u}\right)}\cdot{\sigma}^{-\tfrac{1}{2}},\quad
                           \sigma = \left\Vert \vect{v}\left(\vect{u}\right) \right\Vert_2^2\neq 0.
\end{equation}
From here on, we suppress arguments, such as $\vect{u}$, whenever they are clear from the context.
Further, $\mat{J}_\vect{v}\in\mathbb{R}^{4\times 3}$ is the Jacobian of $\vect{v}$.
Its $j$th column is $\left(\mat{J}_\vect{v} \right)_j=\tfrac{\partial \vect{v}}{\partial u_j}\in\mathbb{R}^{4}$.
We use an analogous notation for the Jacobian of $\vect{f}$.
We write $\mat{I}$ for an identity matrix, $\vect{e}_i$ for the $i$th canonical unit vector, and $\vect{1}$ is a vector whose components are $1$.
The dimensions become apparent from the context.

\subsection{Error Metrics}\label{Sec:ErrorMetrics}
Let $\vect{u}$ be a grid point and $\vect{u}+\varepsilon\cdot\vect{p}$, $\vect{p}\in[0,1]^{3}$,
the parameters of an input unit vector in the grid cell whose lowest corner is $\vect{u}$.
We do not tie the grid to a particular quantizer, so our bounds hold for any placement of the samples.
The encoder replaces them by a corner $\vect{u}+\varepsilon\cdot\vect{b}$ of that cell,
chosen from the candidates $\vect{b}\in\mathcal{B}(\vect{p})\subseteq\{0,1\}^{3}$.
The maximum error of \cref{Eq:QuantizationError} over all inputs is
\begin{equation}
    \Delta Q_\mathrm{exact} = \max_{\vect{u},\vect{p}}\ \min_{\vect{b}\in\mathcal{B}(\vect{p})}
    \underbrace{\angle\left(
        \vect{f}\left(\vect{u}+\varepsilon\cdot\vect{p}\right),\vect{f}\left(\vect{u}+\varepsilon\cdot\vect{b}\right)\right)}_
        {\alpha\left(\vect{u},\vect{p},\vect{b}\right)
        }.
        \label{Eq:QuantErrorGeneral}
\end{equation}
Range quantization considers all cell corners, thus $\mathcal{B}^\mathrm{range}(\vect{p})=\{0,1\}^{3}$.
Domain quantization rounds each parameter, so $\mathcal{B}^\mathrm{domain}(\vect{p})=\{\mathrm{round}(\vect{p})\}$.
For small angles, the geodesic and Euclidean distances converge:
\begin{equation}
    \alpha\left(\vect{u},\vect{p},\vect{b}\right)
    = \left\Vert
        \vect{f}\left(\vect{u}+\varepsilon\cdot\vect{p}\right) -
        \vect{f}\left(\vect{u}+\varepsilon\cdot\vect{b}\right)
      \right\Vert_2
    +\mathcal{O}\left(\varepsilon^2\right).
\end{equation}
Similar to how the metric distortion is computed for mesh parameterization~\cite{Hormann08MPT},
we use the first-order Taylor approximation:
\begin{equation}
    \vect{f}\left(\vect{u}+\varepsilon\cdot\vect{p}\right) - \vect{f}\left(\vect{u}+\varepsilon\cdot\vect{b}\right)=
    \mat{J}_\vect{f}\left(\vect{u}\right)\underbrace{\left(\vect{p}-\vect{b}\right)}_{\vect{y}\in\mathbb{R}^{3}}\cdot\varepsilon
    +\mathcal{O}\left(\varepsilon^2\right).
    \label{Eq:DeltaQGeneral}
\end{equation}
We call $\vect{y}=\vect{p}-\vect{b}$ the rounding error.
Since we seek only first-order approximations and our parameterizations are $\mathcal{C}^2$, Taylor's theorem guarantees that the remainder is $\mathcal{O}\left(\varepsilon^2\right)$, and we drop it.
Squaring does not change the location of the maximum, and
$\left\Vert\mat{J}_\vect{f}\,\vect{y}\cdot\varepsilon\right\Vert_2^2=\vect{y}^\top\,\mat{G}\,\vect{y}\cdot\varepsilon^2$
with
$\mat{G}=\mat{J}_\vect{f}^\top\mat{J}_\vect{f}\in\mathbb{R}^{3\times3}$.
The constant factor $\varepsilon^2$ does not change the location of the maximum, either.
Thus, to first order in $\varepsilon$, our optimization function becomes
\begin{equation}
    Q\left(\vect{u},\vect{y}\right)=\vect{y}^\top\mat{G}(\vect{u})\,\vect{y}.
    \label{Eq:QuantOptFunc}
\end{equation}
We call the first-order maximum error $\Delta Q$.
It satisfies $\Delta Q_\mathrm{exact}=\Delta Q+\mathcal{O}\left(\varepsilon^2\right)$ and can be computed by
\begin{equation}
    \Delta Q = \varepsilon\sqrt{\max_{\vect{u},\vect{p}}\ \min_{\vect{b}\in\mathcal{B}(\vect{p})}\ Q\left(\vect{u},\vect{y}\right)},
    \qquad \vect{y}=\vect{p}-\vect{b}.
    \label{Eq:QuantErrorQuadratic}
\end{equation}

Next, we bring \cref{Eq:QuantOptFunc} into a form from which we can read off the maximum more easily.
To do so, we first compute $\mat{J}_\vect{f}$ from $\mat{J}_\vect{v}$ by applying the product rule to \cref{Eq:DefineVSigma}:
\begin{equation}
    \frac{\partial \vect{f}}{\partial u_j} =
    \frac{\partial}{\partial u_j}\left(\vect{v}\,{\sigma}^{-\tfrac{1}{2}}\right) =
    \frac{\partial \vect{v}}{\partial u_j}\cdot\sigma^{-\tfrac{1}{2}} +
    \vect{v} \cdot \frac{\partial }{\partial u_j}\sigma^{-\tfrac{1}{2}}.
\end{equation}
Since $\sigma=\left\Vert \vect{v} \right\Vert_2^2=\vect{v}^\top\vect{v}$ depends on $\vect{u}$, we need $\tfrac{\partial \sigma}{\partial u_j}=\tfrac{\partial \vect{v}^\top \vect{v}}{\partial u_j}=2\vect{v}^\top\tfrac{\partial \vect{v}}{\partial u_j}$.
Applying the chain rule and inserting this back yields
\begin{equation}
\frac{\partial \vect{f}}{\partial u_j} =
    \sigma^{-\tfrac{1}{2}}\frac{\partial \vect{v}}{\partial u_j} -
    \sigma^{-\tfrac{3}{2}} \vect{v}\left(\vect{v}^\top \frac{\partial \vect{v}}{\partial u_j}\right).
\end{equation}
With some further simplifications and $\left(\mat{J}_\vect{v}\right)_j=\tfrac{\partial \vect{v}}{\partial u_j}$, we get:
\begin{equation}
    \mat{J}_\vect{f}=\frac{1}{\sqrt{\sigma}}
    \mat{F}
    \mat{J}_\vect{v}, \quad  \quad \mat{F}=\left(\mat{I} - \frac{\vect{v}\vect{v}^\top}{\sigma}\right)=\left(\mat{I} - \vect{f}\vect{f}^\top\right).
    \label{Eq:JfFromJv}
\end{equation}
Note that $\mat{F}^\top=\mat{F}\in\mathbb{R}^{4\times 4}$ and, since $\vect{f}^\top\vect{f}=1$, $\mat{F}^\top\mat{F}=\mat{F}^2=\mat{F}$.

With \cref{Eq:JfFromJv} we get
\begin{equation}
    \mat{G}=\mat{J}_\vect{f}^\top\mat{J}_\vect{f}
           = \left(\frac{1}{\sqrt{\sigma}}
             \mat{F}
             \mat{J}_\vect{v}\right)^\top
             \left(\frac{1}{\sqrt{\sigma}}
             \mat{F}
             \mat{J}_\vect{v}\right)
             =\frac{1}{\sigma}
             \mat{J}_\vect{v}^\top \mat{F}^\top\mat{F} \mat{J}_\vect{v}
             =\frac{1}{\sigma}
             \mat{J}_\vect{v}^\top \mat{F} \mat{J}_\vect{v},
\end{equation}
which we then insert into \cref{Eq:QuantOptFunc}:
\begin{align}
Q&= \vect{y}^\top\,
    \mat{G}(\vect{u})\,
    \vect{y}
  =
    \frac{1}{\sigma}\,
    \underbrace{\vect{y}^\top\,\mat{J}_\vect{v}^\top}_{\vect{z}^\top}\,
    \underbrace{\mat{F}}_{\mat{F}^\top\,\mat{F}}\,
    \underbrace{\mat{J}_\vect{v}\,\vect{y}}_{\vect{z}}\\
&=  \frac{1}{\sigma}\,
    \left\Vert {\mat{F}}\,\vect{z}\right\Vert_2^2
    =\frac{1}{\sigma}\,
    \left\Vert \left(\mat{I}-\vect{f}\vect{f}^\top\right)\vect{z}\right\Vert_2^2.
\end{align}
Here, $\vect{z}=\mat{J}_\vect{v}\,\vect{y}$ is the rounding error $\vect{y}$ carried over to $\vect{v}$, i.e., the first-order change of $\vect{v}$, in units of $\varepsilon$.
Simplifying the terms of $\left\Vert \left(\mat{I}-\vect{f}\vect{f}^\top\right)\vect{z}\right\Vert_2^2$ and observing that $\vect{f}^\top\vect{f}=1$ yields
\begin{equation}
    Q
    =\frac{\left\Vert\vect{z}\right\Vert_2^2 - \left(\vect{f}^\top\vect{z}\right)^2}{\sigma}
    =\frac{\left\Vert\vect{z}\right\Vert_2^2}{\sigma} - \frac{\left(\vect{v}^\top\vect{z}\right)^2}{\sigma^2},
    \quad \vect{z}=\mat{J}_\vect{v}\,\vect{y}\in\mathbb{R}^{4}.
    \label{Eq:QuantOptFuncZ}
\end{equation}
We bound $Q$ further
\begin{equation}
    Q
    \leq\frac{\left\Vert\vect{z}\right\Vert_2^2}{\sigma}.
    \label{Eq:QuantOptFuncZBound}
\end{equation}

\subsection{Domain Quantization}
We have $\mathcal{B}^\mathrm{domain}(\vect{p})=\{\mathrm{round}(\vect{p})\}$,
thus $\vect{y}=\vect{p}-\vect{b}\in\left[-\frac{1}{2},\frac{1}{2}\right]^{3}$.
Further, $Q$ is convex in $\vect{y}$, because $\mat{G}$ is positive semidefinite.
Therefore, it is sufficient to consider only $\vect{y}\in\left\{\pm\nicefrac{1}{2}\right\}^{3}$ in \cref{Eq:DeltaQGeneral}.
\begin{equation}
    \Delta Q^\mathrm{domain} = \varepsilon\sqrt{\max_{\vect{u},\vect{y}\in\left\{\pm\nicefrac{1}{2}\right\}^{3}}\ Q\left(\vect{u},\vect{y}\right)}\label{Eq:DomainErrorQuadratic}.
\end{equation}

\subsubsection{$\lOne$-Parameterization}\label{Sec:LOneDomainProof}
Since we assume $u_i\geq0$, $\vect{v}=\left[u_1,u_2,u_3,1-u_1-u_2-u_3\right]^\top$ for $\lOne$.
To bound the denominator of \cref{Eq:QuantOptFuncZBound}, we use that $v_k$ sum to 1 and the Cauchy--Schwarz inequality
\begin{align}
    1
    = \left(\sum_{k=1}^{4}v_k\right)^2
    = \left(\vect{1}^\top\vect{v}\right)^2
    \leq \left\Vert \vect{1} \right\Vert_2^2 \cdot \left\Vert\vect{v}\right\Vert_2^2
    = 4\sigma.
\end{align}
Thus, $\sigma\geq \frac{1}{4}$.
To find the location of the maxima, equality must hold.
That is the case, if $\vect{v}$ is a multiple of $\vect{1}$, i.e., all components of $\vect{v}$ must be equal.
Since they sum to $1$, $v_i=\frac{1}{4}$.

The Jacobian of $\vect{v}$ and the offset $\vect{z}$ are
\begin{equation}
    \mat{J}_\vect{v}=
    \begin{bmatrix}
         1 &  0 &  0\\
         0 &  1 &  0\\
         0 &  0 &  1\\
        -1 & -1 & -1
    \end{bmatrix},\quad
    \vect{z}=\mat{J}_\vect{v}\vect{y}=
    \begin{bmatrix}
        y_1\\
        y_2\\
        y_3\\
        -y_1-y_2-y_3
    \end{bmatrix}.
\end{equation}
Note that the components of $\vect{z}$ sum to $0$ and $\vect{v}^\top\vect{z}=0$ for $v_i=\frac{1}{4}$.
This makes the bound tight for \cref{Eq:QuantOptFuncZBound}.

Next, we bound the numerator of \cref{Eq:QuantOptFuncZBound} with
\begin{align}
\Vert\vect{z}\Vert_2^2
    = y_1^2+y_2^2+y_3^2 + \left(y_1+y_2+y_3\right)^2
    \leq \frac{3}{4} + \left(\frac{3}{2}\right)^2
    = 3.
    \label{Eq:LOneZNorm}
\end{align}
Equality holds for $y_k=\frac{1}{2}$.

Putting the two bounds together into \cref{Eq:QuantOptFuncZBound}, we get
\begin{align}
Q
    \leq\frac{\left\Vert\vect{z}\right\Vert_2^2}{\sigma}
    \leq\frac{3}{\frac{1}{4}}
    = 12.
\end{align}
With \cref{Eq:DomainErrorQuadratic}, we obtain the maximum error and its location:
\begin{equation}
    \label{Eq:LOneDomainResult}
    \boxed{%
    \begin{aligned}
        \Delta Q^{\lOne,\mathrm{domain}} &=2\sqrt{3}\cdot\varepsilon,\\
        \vect{u}^{\lOne,\mathrm{domain}} &=\left[\frac{1}{4}, \frac{1}{4}, \frac{1}{4}\right]^\top.
    \end{aligned}}
\end{equation}

\subsubsection{$\lInf$-Parameterization}
For $\lInf$, $\sigma\geq 1$ (cf.\ \cref{Eq:DefineVSigma}), because at least one component of $\vect{v}=\left[u_1,u_2,u_3,1\right]^\top$ has magnitude one.
This provides a lower bound for the denominator of \cref{Eq:QuantOptFuncZBound}.
For its numerator, we compute
\begin{equation}
    \mat{J}_\vect{v}=
    \begin{bmatrix}
        1 & 0 & 0\\
        0 & 1 & 0\\
        0 & 0 & 1\\
        0 & 0 & 0
    \end{bmatrix},\quad
    \vect{z}=\mat{J}_\vect{v}\vect{y}=
    \begin{bmatrix}
        y_1\\
        y_2\\
        y_3\\
        0
    \end{bmatrix}.
\end{equation}
This lets us further bound \cref{Eq:QuantOptFuncZBound}:
\begin{equation}
    Q\leq\frac{\left\Vert\vect{z}\right\Vert_2^2}{\sigma}\leq\frac{y_1^2+y_2^2+y_3^2}{1}
    =
    \frac{3}{4}.
\end{equation}
The bound becomes tight if $\sigma = 1 = \left\Vert\vect{v}\right\Vert_2^2=u_1^2+u_2^2+u_3^2+1$.
This can only hold if $\vect{u}=\vect{0}$.
Then the subtracted term of \cref{Eq:QuantOptFuncZ} vanishes, because
$\vect{v}^\top\vect{z}=\left[0,0,0,1\right]\left[y_1,y_2,y_3,0\right]^\top=0$.
With \cref{Eq:DomainErrorQuadratic}, we obtain the maximum error and its location:
\begin{equation}
    \boxed{%
    \begin{aligned}
        \Delta Q^{\lInf,\mathrm{domain}} &=\frac{\sqrt{3}}{2}\cdot\varepsilon,\\
        u^{\lInf,\mathrm{domain}}      &=\left[0,0,0\right]^\top.
    \end{aligned}}
    \label{Eq:LInfDomainResult}
\end{equation}

\subsubsection{$\lInfTan$-Parameterization}\label{Sec:LInfTanDomainProof}
Here, we get $v_i=\tan{\left(\frac{\pi}{4}u_i\right)}$ for $1\leq i \leq 3$ and $v_4=1$.
So $\sigma=\Vert \vect{v} \Vert_2^2 = 1+\rho $ where $\rho=v_1^2+v_2^2+v_3^2$.
Since the derivative of $\tan{\left(\frac{\pi}{4}u_i\right)}$ is $\frac{\pi}{4}\left(1+v_i^2\right)$,
\begin{equation}
    \mat{J}_\vect{v}=\frac{\pi}{4}
    \begin{bmatrix}
        1+v_1^2 & 0 & 0\\
        0 & 1+v_2^2 & 0\\
        0 & 0 & 1+v_3^2\\
        0 & 0 & 0
    \end{bmatrix},\quad
    \vect{z}=\frac{\pi}{4}
    \begin{bmatrix}
        \left(1+v_1^2\right)y_1\\
        \left(1+v_2^2\right)y_2\\
        \left(1+v_3^2\right)y_3\\
        0
    \end{bmatrix}.
\end{equation}
Since $y_i^2=\nicefrac{1}{4}$,
\begin{equation}
\Vert \vect{z} \Vert^2_2 = \frac{\pi^2}{64}\sum_{k=1}^{3}\left(1+v_k^2\right)^2
= \frac{\pi^2}{64}\sum_{k=1}^{3}\left(1+2\,v_k^2+v_k^4\right).
\end{equation}
Because $0\leq u_i\leq 1$, $0 \leq v_i \leq 1$ and hence $v_i^4 \leq v_i^2$:
\begin{equation}
\Vert \vect{z} \Vert^2_2 \leq
 \frac{\pi^2}{64}\sum_{k=1}^{3}\left(1+3\,v_k^2\right)
= \frac{\pi^2}{64}\left(3 + 3\,\rho\right)
= \frac{3\,\pi^2}{64}\,\sigma.
\end{equation}
This lets us bound \cref{Eq:QuantOptFuncZBound}
\begin{equation}
    Q\leq\frac{\Vert \vect{z} \Vert^2_2}{\sigma}\leq\frac{3\,\pi^2}{64}.
\end{equation}
At $\vect{u}=\vect{0}$, all $v_i=0$ and $\rho=0$.
Thus, every inequality above becomes an equality and the subtracted term of \cref{Eq:QuantOptFuncZ} vanishes.
Inserting into \cref{Eq:DomainErrorQuadratic} gives us:
\begin{equation}
    \label{Eq:LInfTanDomainResult}
    \boxed{%
    \begin{aligned}
        \Delta Q^{\lInfTan,\mathrm{domain}} &=\frac{\pi\sqrt{3}}{8}\cdot\varepsilon,\\
        u^{\lInfTan,\mathrm{domain}}      &=\left[0,0,0\right]^\top
    \end{aligned}}
\end{equation}

\subsubsection{$\lTwoI$-Parameterization}\label{Sec:LTwoIDomainProof}
Here, $\vect{v}=\left[u_1,u_2,u_3,v_4\right]^\top$ with $v_4^2=1-u_1^2-u_2^2-u_3^2$.
Note that $\lTwoI$ permutes the component such that $0\leq u_i\leq v_4$.
We compute
\begin{equation}
    \mat{J}_\vect{v}=
    \begin{bmatrix}
        1 & 0 & 0\\
        0 & 1 & 0\\
        0 & 0 & 1\\
        -\frac{u_1}{v_4} & -\frac{u_2}{v_4} & -\frac{u_3}{v_4}
    \end{bmatrix},\quad
    \vect{z}=\mat{J}_\vect{v}\vect{y}=
    \begin{bmatrix}
        y_1\\
        y_2\\
        y_3\\
        -\frac{\vect{u}^\top\vect{y}}{v_4}
    \end{bmatrix}.
\end{equation}
This parameterization already produces normalized vectors, thus $\sigma=\vect{v}^\top\vect{v}=1$ and the denominators in \cref{Eq:QuantOptFuncZ} become $1$.
Further, the subtracted term vanishes too, since
\begin{equation}
    \vect{v}^\top\vect{z}
    = u_1y_1+u_2y_2+u_3y_3 - v_4\cdot \frac{\vect{u}^\top\vect{y}}{v_4}
    = 0.
\end{equation}
Hence, for every offset $\vect{y}$,
\begin{equation}
    Q = \Vert \vect{z} \Vert_2^2
    =y_1^2+y_2^2+y_3^2 + \frac{\left(\vect{u}^\top\vect{y}\right)^2}{v_4^2}.
    \label{Eq:LTwoIQ}
\end{equation}
For domain quantization, $y_i^2=\nicefrac{1}{4}$, so $Q=\frac{3}{4}+\frac{\left(\vect{u}^\top\vect{y}\right)^2}{v_4^2}$.
To bound the second term we use $\vert y_i\vert =\nicefrac{1}{2}$ and $0\leq u_i\leq v_4$:
\begin{equation}
     {\left(\vect{u}^\top\vect{y}\right)^2}
     =    \left(u_1y_1+u_2y_2+u_3y_3\right)^2
     \leq \left(3\cdot v_4\cdot \frac{1}{2}\right)^2
     =    \frac{9\,v_4^2}{4},
     \label{Eq:L2InfBounduTy}
\end{equation}
and thus:
\begin{equation}
    Q \leq \frac{3}{4} + \frac{\frac{9\,v_4^2}{4}}{v_4^2}
      = \frac{3}{4} + \frac{9}{4}
      = 3.
\end{equation}
For the bound on $Q$, \cref{Eq:L2InfBounduTy} must be attained.
This is the case if $y_i=\nicefrac{1}{2}$ and $u_i=v_4$.
That means $\left\Vert\vect{v}\right\Vert_2^2=4\,v_4^2=1$ and therefore $u_i = \frac{1}{2}$.
Since $\lTwoI$ scales its parameters by $\sqrt{2}$ before quantization, the grid spacing in $\vect{u}$ is $\frac{\sqrt{2}}{2}\varepsilon$ instead of $\varepsilon$.
Inserting into \cref{Eq:DomainErrorQuadratic} with this spacing gives $\Delta Q=\frac{\sqrt{2}}{2}\varepsilon\cdot\sqrt{3}$:
\begin{equation}
    \boxed{
        \begin{aligned}
            \Delta Q^{\lTwoI,\mathrm{domain}} &=\frac{\sqrt{6}}{2}\cdot\varepsilon,\\
            u^{\lTwoI,\mathrm{domain}}      &=\left[\frac{1}{2},\frac{1}{2},\frac{1}{2}\right]^\top
        \end{aligned}}
    \label{Eq:LTwoIDomainResult}
\end{equation}

\subsubsection{Exp-Parameterization}\label{Sec:ExpDomainProof}
For the chart $c=0$, $\vect{v}=\left[\sin\left(\upsilon\right)\frac{\vect{u}^\top}{\Vert\vect{u}\Vert_2},\cos\left(\upsilon\right)\right]^\top$ with $\upsilon=\frac{\pi}{2}\Vert\vect{u}\Vert_2$.
Here $\vect{f}=\vect{v}$ because this parameterization already produces a unit vector.
Therefore, $\sigma=\vect{v}^\top\vect{v}=1$.
With $s=\frac{\sin\upsilon}{\Vert\vect{u}\Vert_2}$, $\uvec{u}=\frac{\vect{u}}{\Vert\vect{u}\Vert_2}$, and $\frac{\partial\Vert\vect{u}\Vert_2}{\partial u_j}=\ucomp{u}_j$, differentiating $\vect{v}$ yields
\begin{equation}
    \mat{J}_\vect{v}=
    \begin{bmatrix}
        s\,\mat{I}+\left(\frac{\pi}{2}\cos\upsilon-s\right)\uvec{u}\uvec{u}^\top\\[2pt]
        -\frac{\pi}{2}\sin\left(\upsilon\right)\uvec{u}^\top
    \end{bmatrix}.
\end{equation}
Multiplying with $\vect{y}$ gives
\begin{equation}
    \vect{z}=\mat{J}_\vect{v}\vect{y}=
    \begin{bmatrix}
        s\,\vect{y}+\left(\frac{\pi}{2}\cos\upsilon-s\right)\uvec{u}\left(\uvec{u}^\top\vect{y}\right)\\[2pt]
        -\frac{\pi}{2}\sin\left(\upsilon\right)\left(\uvec{u}^\top\vect{y}\right)
    \end{bmatrix}.
\end{equation}
The subtracted term of \cref{Eq:QuantOptFuncZ} vanishes, since $\vect{v}=\left[\sin\left(\upsilon\right)\uvec{u}^\top,\cos\left(\upsilon\right)\right]^\top$ and $\uvec{u}^\top\uvec{u}=1$ give
\begin{equation}
    \begin{aligned}
        \vect{v}^\top\vect{z}
        &=\sin\upsilon\left(s+\tfrac{\pi}{2}\cos\upsilon-s\right)\left(\uvec{u}^\top\vect{y}\right)
         -\cos\upsilon\,\tfrac{\pi}{2}\sin\upsilon\left(\uvec{u}^\top\vect{y}\right)\\
        &=\tfrac{\pi}{2}\sin\upsilon\cos\upsilon\left(\uvec{u}^\top\vect{y}\right)
         -\tfrac{\pi}{2}\sin\upsilon\cos\upsilon\left(\uvec{u}^\top\vect{y}\right)
        =0.
    \end{aligned}
\end{equation}
Hence, $Q=\Vert\vect{z}\Vert_2^2$.
We expand $\Vert\vect{z}\Vert_2^2$ block by block.
The upper block is a sum of $s\,\vect{y}$ and a multiple of $\uvec{u}$.
Its squared norm consists of the squared norms of both summands and twice their dot product.
With $\uvec{u}^\top\uvec{u}=1$, these are
\begin{equation}
    \underbrace{s^2\Vert\vect{y}\Vert_2^2}_{\text{first summand}}
    +\underbrace{2s\left(\tfrac{\pi}{2}\cos\upsilon-s\right)\left(\uvec{u}^\top\vect{y}\right)^2}_{\text{dot product}}
    +\underbrace{\left(\tfrac{\pi}{2}\cos\upsilon-s\right)^2\left(\uvec{u}^\top\vect{y}\right)^2}_{\text{second summand}}.
\end{equation}
The last two terms share the factor $\left(\uvec{u}^\top\vect{y}\right)^2$, and their coefficients add up to
\begin{equation}
    \begin{aligned}
        &2s\left(\tfrac{\pi}{2}\cos\upsilon-s\right)+\left(\tfrac{\pi}{2}\cos\upsilon-s\right)^2\\
        &\quad=\left(\tfrac{\pi}{2}\cos\upsilon-s\right)\left(\tfrac{\pi}{2}\cos\upsilon+s\right)
        =\tfrac{\pi^2}{4}\cos^2\upsilon-s^2.
    \end{aligned}
\end{equation}
The lower block contributes $\frac{\pi^2}{4}\sin^2\left(\upsilon\right)\left(\uvec{u}^\top\vect{y}\right)^2$.
Adding both blocks and using $\cos^2\upsilon+\sin^2\upsilon=1$ yields
\begin{equation}
    Q=\Vert\vect{z}\Vert_2^2=s^2\Vert\vect{y}\Vert_2^2+\left(\frac{\pi^2}{4}-s^2\right)\left(\uvec{u}^\top\vect{y}\right)^2.
\end{equation}
Since $\sin\upsilon\leq\upsilon$, $s\leq\frac{\pi}{2}$, and since $\left(\uvec{u}^\top\vect{y}\right)^2\leq\Vert\vect{y}\Vert_2^2$,
\begin{equation}
    Q\leq\frac{\pi^2}{4}\Vert\vect{y}\Vert_2^2=\frac{\pi^2}{4}\cdot\frac{3}{4}=\frac{3\,\pi^2}{16}.
\end{equation}
At $\vect{u}=\vect{0}$, $s=\frac{\pi}{2}$ and equality holds for every $\vect{y}$.
Inserting into \cref{Eq:DomainErrorQuadratic} gives:
\begin{equation}
    \boxed{
        \begin{aligned}
            \Delta Q^{\mathrm{Exp},\mathrm{domain}} &=\frac{\pi\sqrt{3}}{4}\cdot\varepsilon,\\
            \vect{u}^{\mathrm{Exp},\mathrm{domain}}      &=\left[0,0,0\right]^\top.
        \end{aligned}}
    \label{Eq:ExpDomainResult}
\end{equation}

\subsection{Range Quantization}\label{Sec:RangeQuantizationProof}
Range quantization picks the cell corner that minimizes the angle to the input unit vector.
Since $\mathcal{B}^\mathrm{domain}\subseteq\mathcal{B}^\mathrm{range}$, it can always pick the same corner as domain quantization, so $\Delta Q^\mathrm{range}\leq\Delta Q^\mathrm{domain}$.

The worst input point for domain quantization is the cell center.
At the cell center, the offset to every corner is $\vect{y}\in\left\{\pm\frac{1}{2}\right\}^{3}$.
If $\mat{G}$ is diagonal, $Q(\vect{u},\vect{y})=\vect{y}^\top \mat{G} \vect{y}=G_{1,1}y_1^2+G_{2,2}y_2^2+G_{3,3}y_3^2=\frac{1}{4}\left(G_{1,1}+G_{2,2}+G_{3,3}\right)$ takes the same value for every choice of $\vect{y}$.
For this input, range quantization cannot improve on domain quantization.
Together with $\Delta Q^\mathrm{range}\leq\Delta Q^\mathrm{domain}$, both maximum errors coincide.
This is the case for $\lInf$, $\lInfTan$, and Exp, because at their domain maximizer $\vect{u}=\vect{0}$, $\mat{G}$ is diagonal.

\subsubsection{$\lOne$-Parameterization}\label{Sec:LOneRangeProof}
The identity in \cref{Eq:LOneZNorm} holds for every corner:
\begin{equation}
\Vert\vect{z}\Vert_2^2=\Vert\vect{y}\Vert_2^2+\left(\vect{1}^\top\vect{y}\right)^2.
\end{equation}
Range quantization picks the best of all corners, so its error is at most that of any particular corner.
For domain quantization, the second term reaches $\left(\frac{3}{2}\right)^2$.
We now construct a corner whose second term is at most $\frac{1}{4}$, while its first term stays at most $\frac{3}{4}$.

We find this corner as follows:
Recall from \cref{Eq:QuantErrorGeneral,Eq:QuantErrorQuadratic} that $\vect{p}\in[0,1]^3$ is the position of the input in its cell, $\vect{b}\in\{0,1\}^3$ is a corner, and $\vect{y}=\vect{p}-\vect{b}$.
Let $P=\vect{1}^\top\vect{p}\in\left[0,3\right]$ and $m=\mathrm{round}\left(P\right)\in\{0,1,2,3\}$.
We obtain $\vect{b}$ from $\vect{p}$ by setting its $m$ largest components to $1$ and the others to $0$.
Since the components of $\vect{b}$ sum to $m$, $\vect{1}^\top\vect{y}=P-m$, so 
\begin{equation}
    \left(\vect{1}^\top\vect{y}\right)^2=(P-m)^2\leq\left(\frac{1}{2}\right)^2=\frac{1}{4}.
\end{equation}

Further, $\Vert\vect{y}\Vert_2^2\leq\frac{3}{4}$ still holds:
For $m=0$, $\vect{b}=\vect{0}$, therefore $\vect{y}=\vect{p}\geq\vect{0}$.
Because $m=\mathrm{round}(P)=0$, $P=p_1+p_2+p_3\leq\frac{1}{2}$.
Therefore,
\begin{equation}
    \Vert\vect{y}\Vert_2^2=\Vert\vect{p}\Vert_2^2=p_1^2+p_2^2+p_3^2\leq(p_1+p_2+p_3)^2=P^2\leq\frac{1}{4}.
\end{equation}

For $m=1$, i.e., $\frac{1}{2}\leq P\leq\frac{3}{2}$, let $p_k$ be the largest component.
Further, $b_k=1$ and the other $b_i$ are zero.
\begin{align}
    \Vert\vect{y}\Vert_2^2=\Vert\vect{p}-\vect{b}\Vert_2^2    
    &=\left(p_k-1\right)^2+\sum_{i\neq k}(p_i-0)^2\\
    &= \left(p_k-1\right)^2+\sum_{i\neq k}p_i^2.
\end{align} 
Because $0\leq p_i\leq p_k$ for $i\neq k$, we have $p_i^2\leq p_k\,p_i$:
\begin{equation}
    \Vert\vect{y}\Vert_2^2 \leq \left(p_k-1\right)^2+\sum_{i\neq k}p_k\cdot p_i = \left(p_k-1\right)^2 + p_k\cdot\sum_{i\neq k} p_i.
\end{equation}
Further, $\sum_{i\neq k} p_i = P - p_k$.
\begin{align}
    \Vert\vect{y}\Vert_2^2
    &\leq\left(p_k-1\right)^2+p_k\left(P-p_k\right)
    = p_k^2-2\cdot p_k + 1 +p_k\,P-p_k^2\\
    &= -2\cdot p_k + 1 +P\,p_k
    =1-p_k\left(2-P\right).
\end{align}
Because $2-P>0$ and $p_k$, as the largest component, is at least the average, $p_k\geq\frac{P}{3}$, we can replace $p_k$ by $\frac{P}{3}$.
Since $m=1$ means $\vert P-1\vert\leq\frac{1}{2}$, we get
\begin{equation}
    \Vert\vect{y}\Vert_2^2\leq1-\frac{P}{3}\left(2-P\right)=\frac{2+\left(P-1\right)^2}{3}\leq\frac{2+\frac{1}{4}}{3}=\frac{3}{4}.
\end{equation}
The cases $m=3$ and $m=2$ follow from $m=0$ and $m=1$ by substituting $\vect{p}$ with $\vect{1}-\vect{p}$, which negates $\vect{y}$.

Hence, $\Vert\vect{z}\Vert_2^2\leq\frac{3}{4}+\frac{1}{4}=1$, and with $\sigma\geq\frac{1}{4}$ from the domain case, $Q\leq4$.

Equality holds for $\vect{p}=\frac{1}{2}\vect{1}$, where every corner has $\Vert\vect{z}\Vert_2^2\geq1$, at $\vect{u}=\frac{1}{4}\vect{1}$, where $\sigma=\frac{1}{4}$ and $\vect{v}^\top\vect{z}=0$.
Inserting into \cref{Eq:DomainErrorQuadratic} yields:
\begin{equation}
    \boxed{
        \begin{aligned}
            \Delta Q^{\lOne,\mathrm{range}} &= 2\,\varepsilon,\\
            \vect{u}^{\lOne,\mathrm{range}} &=\left[\frac{1}{4},\frac{1}{4},\frac{1}{4}\right]^\top.
        \end{aligned}}
\end{equation}

\subsubsection{$\lTwoI$-Parameterization}\label{Sec:LTwoIRangeProof}
The offset $\vect{z}$ and \cref{Eq:LTwoIQ} from \cref{Sec:LTwoIDomainProof} hold for every offset $\vect{y}$.
With $\pmb{\omega}=\frac{\vect{u}}{v_4}\in[0,1]^3$, which follows from $0\leq u_i\leq v_4$, they read
\begin{equation}
    \vect{z}=\begin{bmatrix}\vect{y}\\-\pmb{\omega}^\top\vect{y}\end{bmatrix},\qquad
    Q=\Vert\vect{z}\Vert_2^2=\Vert\vect{y}\Vert_2^2+\left(\pmb{\omega}^\top\vect{y}\right)^2.
    \label{Eq:LTwoIRangeQ}
\end{equation}
Range quantization picks the best corner.
Therefore, it suffices to show that the average of $Q$ over a \emph{random corner} is at most $1$.

Domain quantization rounds each component on its own, $b_k=\left\lfloor p_k+\frac{1}{2}\right\rfloor$.
In the worst case, all rounding errors $y_k$ have the same sign and add up in $\pmb{\omega}^\top\vect{y}$, so the second term of $Q$ reaches $\left(\frac{3}{2}\right)^2$.

To obtain a random corner, we replace the fixed threshold $\frac{1}{2}$ by a random $\xi$, drawn uniformly from $[0,1)$, i.e., we dither:
\begin{equation}
    b_k=\left\lfloor p_k+\xi\right\rfloor.
\end{equation}
It rounds $p_k$ up with probability $p_k$, so it is correct on average.
However, the rounding errors of the components can still add up in $\pmb{\omega}^\top\vect{y}$.
We avoid this by rounding \emph{running sums} instead of components.
First, we renumber the axes such that $\omega_1\geq \omega_2\geq \omega_3$, which renumbers the components of $\vect{p}$, $\vect{b}$, and $\vect{y}$ alike and leaves $Q$ unchanged.
The running sums are
\begin{equation}
    s_0=0,\quad s_1=p_1,\quad s_2=p_1+p_2,\quad s_3=p_1+p_2+p_3.
\end{equation}
Rounding $s_k$ instead of the components keeps the sum $y_1+\dots+y_k$ of the first $k$ rounding errors small.
With decreasing weights, $\pmb{\omega}^\top\vect{y}$ can be expressed as a convex combination of the errors of the running sums, with coefficients from $\pmb{\omega}$, as we show below.
We dither all running sums with the same $\xi$ and obtain the components of $\vect{b}$ as differences:
\begin{equation}
    T_k=\left\lfloor s_k+\xi\right\rfloor,\qquad b_k=T_k-T_{k-1}.
\end{equation}
Since $T_0=\lfloor \xi\rfloor=0$, inverting the differences gives
\begin{equation}
    T_1=b_1,\quad T_2=b_1+b_2,\quad T_3=b_1+b_2+b_3,
\end{equation}
analogous to the running sums $s_k$ of $\vect{p}$.
Subtracting these from the running sums $s_k$ of $\vect{p}$ yields the running sums of the rounding errors $\vect{y}=\vect{p}-\vect{b}$ from \cref{Eq:DeltaQGeneral}:
\begin{equation}
    \begin{aligned}
        s_1-T_1&=p_1-b_1=y_1,\\
        s_2-T_2&=\left(p_1+p_2\right)-\left(b_1+b_2\right)=y_1+y_2,\\
        s_3-T_3&=\left(p_1+p_2+p_3\right)-\left(b_1+b_2+b_3\right)=y_1+y_2+y_3.
    \end{aligned}
    \label{Eq:RunningErrors}
\end{equation}
Since $T_k$ is $s_k$ rounded to a neighboring integer, all three stay below $1$ in magnitude: $\left\vert y_1\right\vert<1$, $\left\vert y_1+y_2\right\vert<1$, and $\left\vert y_1+y_2+y_3\right\vert<1$.
Since $0\leq p_k\leq1$, we have $b_k\in\{0,1\}$, so $\vect{b}$ is a corner.

Each rounded running sum is $T_k=\lfloor s_k\rfloor+1$ exactly for $\xi\geq1-r_k$, i.e., with probability equal to the fractional part $r_k=s_k-\lfloor s_k\rfloor$, and $T_k=\lfloor s_k\rfloor$ otherwise.
Hence,
\begin{subequations}\label{Eq:DitherMoments}
\begin{align}
    \mathbb{E}\left[T_k\right]&=\left(\lfloor s_k\rfloor+1\right)r_k+\lfloor s_k\rfloor\left(1-r_k\right)=s_k,\label{Eq:DitherMean}\\
    \mathbb{E}\left[b_k\right]&=\mathbb{E}\left[T_k\right]-\mathbb{E}\left[T_{k-1}\right]=s_k-s_{k-1}=p_k.\label{Eq:DitherB}
\end{align}
\end{subequations}
\Cref{Eq:DitherMean} gives \cref{Eq:DitherB}, which we need for the first term of $Q$:
Since $b_k\in\{0,1\}$ has the expected value $p_k$, it is $1$ with probability $p_k$, so $\left(p_k-b_k\right)^2$ is $\left(1-p_k\right)^2$ with probability $p_k$ and $p_k^2$ otherwise.
Because the expected value of a sum is the sum of the expected values, and $p_k\left(1-p_k\right)\leq\frac{1}{4}$,
\begin{equation}
    \begin{aligned}
        \mathbb{E}\left[\Vert\vect{y}\Vert_2^2\right]
        &=\mathbb{E}\left[\Vert\vect{p}-\vect{b}\Vert_2^2\right]
         =\sum_{k=1}^{3}\mathbb{E}\left[\left(p_k-b_k\right)^2\right]\\
        &=\sum_{k=1}^{3}\left(p_k\left(1-p_k\right)^2+\left(1-p_k\right)p_k^2\right)\\
        &=\sum_{k=1}^{3} p_k\left(1-p_k\right)\leq\frac{3}{4}.
    \end{aligned}
\end{equation}
For the second term of $Q$, we compute $\mathbb{E}\left[\left(\pmb{\omega}^\top\vect{y}\right)^2\right]$.
Therefore, we need the squared errors $d_k^2$ of the rounded running sums, where $d_k=s_k-T_k$ is $r_k-1$ when rounding up and $r_k$ when rounding down:
\begin{equation}
    \mathbb{E}\left[d_k^2\right]=r_k\left(1-r_k\right)^2+\left(1-r_k\right)r_k^2=r_k\left(1-r_k\right)\leq\frac{1}{4}.
    \label{Eq:DitherVar}
\end{equation}
Further, let $\lambda_0=1-\omega_1$ and $\lambda_k=\omega_k-\omega_{k+1}$ for $k=1,2,3$ with $\omega_4=0$.
Since the weights decrease, all $\lambda_k\geq0$, and they sum to $1$.
By \cref{Eq:RunningErrors}, $y_1=d_1$, $y_2=d_2-d_1$, and $y_3=d_3-d_2$.
With $d_0=0$,
\begin{equation}
    \begin{aligned}
        \pmb{\omega}^\top\vect{y}
        &=\omega_1d_1+\omega_2\left(d_2-d_1\right)+\omega_3\left(d_3-d_2\right)\\
        &=\left(\omega_1-\omega_2\right)d_1+\left(\omega_2-\omega_3\right)d_2+\omega_3d_3=\sum_{k=0}^{3}\lambda_k d_k
    \end{aligned}
\end{equation}
is a convex combination.
Since squaring is convex, Jensen's inequality gives
\begin{equation}
    \left(\pmb{\omega}^\top\vect{y}\right)^2=\left(\sum_{k=0}^{3} \lambda_k d_k\right)^2\leq\sum_{k=0}^{3} \lambda_k d_k^2.
\end{equation}
Taking the expected value and bounding each $\mathbb{E}\left[d_k^2\right]$ with \cref{Eq:DitherVar} gives
\begin{equation}
    \mathbb{E}\left[\left(\pmb{\omega}^\top\vect{y}\right)^2\right]
    \leq\sum_{k=0}^{3} \lambda_k\,\mathbb{E}\left[d_k^2\right]\leq\frac{1}{4}\sum_{k=0}^{3} \lambda_k=\frac{1}{4},
\end{equation}
since the $\lambda_k$ are nonnegative and sum to $1$.
Thus, $\mathbb{E}\left[Q\right]\leq\frac{3}{4}+\frac{1}{4}=1$.
If every corner had $Q>1$, the average would exceed $1$.
Hence, the best corner has $Q\leq1$.

At the domain boundary, a coordinate has only one candidate, e.g., $u_1=\frac{\sqrt{2}}{2}$.
There, $u_i\leq v_4$ forces $u_2=u_3=0$, so $\pmb{\omega}=\vect{e}_1$ and $Q=2y_1^2+y_2^2+y_3^2\leq1$ with each coordinate rounded to its nearest candidate.

With the grid spacing $\frac{\sqrt{2}}{2}\varepsilon$ of $\lTwoI$, the error is at most $\frac{\sqrt{2}}{2}\varepsilon$.
Equality holds for $\vect{u}=\left[\frac{\sqrt{2}}{2},0,0\right]^\top$, i.e., $\pmb{\omega}=\vect{e}_1$, and $\vect{p}=\frac{1}{2}\vect{1}$, where every corner has $Q=2y_1^2+y_2^2+y_3^2=1$:
\begin{equation}
    \boxed{
        \begin{aligned}
            \Delta Q^{\lTwoI,\mathrm{range}} &= \frac{\sqrt{2}}{2}\cdot\varepsilon,\\
            \vect{u}^{\lTwoI,\mathrm{range}} &=\left[\frac{\sqrt{2}}{2},0,0\right]^\top.
        \end{aligned}}
    \label{Eq:LTwoIRangeResult}
\end{equation}

\section{Error Analysis of the $\lTwo$-Parameterization }\label{Sec:LTwoProof}
Next, we analyze the $\lTwo$-parameterization, which reconstructs $x_4=\sqrt{1-\Vert\vect{u}\Vert_2^2}$.
To avoid negative arguments of the square root, $\vect{u}$ is projected onto the unit ball, $\Vert\vect{u}\Vert_2\leq1$.
In expressions such as ${1-\Vert\vect{u}\Vert_2^2}$, we assume that this projection has implicitly happened.
We consider only one chart and, on that chart, only non-negative parameters, $u_i\geq0$; the other cases follow from symmetry.

Since the Jacobian of $\lTwo$-parameterization is unbounded where $x_4\rightarrow 0$, the Jacobian-based approach of \cref{Sec:ErrorAnalysisProof} fails for $\lTwo$.
For $s\geq 0$, $t\geq 0$
\begin{equation}
    \left(\sqrt{s} - \sqrt{t}\right)^2 \leq \vert s - t \vert\label{Eq:ST}
\end{equation}   
holds because
\begin{align}
    \left(\sqrt{s} - \sqrt{t}\right)^2 &= s + t - 2\sqrt{st}
    \leq s + t - 2\sqrt{\min\left(s,t\right)^2}\\
                 &= s + t - 2\min\left(s,t\right)
                 = \left\vert s - t \right\vert.
\end{align}

Quantization yields the parameters $\tilde{\vect{u}}$ and the reconstructed component
$\tilde{x}_4=\sqrt{1-\Vert\tilde{\vect{u}}\Vert^2_2}$.
We get $\Delta x_4 = \vert \tilde{x}_4 - x_4 \vert$ from \cref{Eq:ST}:
\begin{equation}
    \Delta x_4^2 = \vert x_4 - \tilde{x}_4  \vert^2
                 = \left(\sqrt{\underbrace{1 - \Vert\vect{u}\Vert_2^2}_{s}} - \sqrt{\underbrace{1-\Vert\tilde{\vect{u}}\Vert_2^2}_{t}}\right)^2
                 \leq \left\vert s - t \right\vert.\label{Eq:L2BoundX4}
\end{equation}
Equality holds if $\min\left(s,t\right)=0$, i.e., if $\vect{u}$ or $\tilde{\vect{u}}$ lies on the equator, where $\Vert\vect{u}\Vert_2=1$ or $\Vert\tilde{\vect{u}}\Vert_2=1$.

\subsection{Domain Quantization}
We further bound $\Delta x_4^2$:
\begin{align}
    \Delta x_4^2&\leq\left\vert s - t \right\vert =
            \left\vert \left(1 - \Vert\vect{u}\Vert^2_2\right) -
                       \left(1-\Vert\tilde{\vect{u}}\Vert^2_2\right)
            \right\vert\\
                &= \left\vert \Vert\tilde{\vect{u}}\Vert^2_2 - \Vert\vect{u}\Vert^2_2\right\vert
                =  \left\vert\left(\tilde{\vect{u}} - \vect{u}\right)^\top\left(\tilde{\vect{u}} + \vect{u}\right)\right\vert
                \leq \left\Vert\tilde{\vect{u}} - \vect{u}\right\Vert_2\left\Vert\tilde{\vect{u}} + \vect{u}\right\Vert_2.
\end{align}
Next, we bound the two factors separately.
A grid cell has a diagonal of length $\sqrt{3}\,\varepsilon$.
Since rounding picks the nearest grid point, $\tilde{\vect{u}}$ and $\vect{u}$ are at most half a cell diagonal apart, so $\Vert\tilde{\vect{u}} - \vect{u}\Vert_2 \leq \frac{\sqrt{3}}{2}\varepsilon$.
Note that this also holds if $\tilde{\vect{u}}$ is projected, since projecting onto the unit ball never moves a point away from $\vect{u}$, which lies inside it.
Further, $\Vert \vect{u} \Vert_2 \leq 1$ and $\Vert \tilde{\vect{u}} \Vert_2 \leq 1$, hence $\Vert\tilde{\vect{u}} + \vect{u}\Vert_2 \leq 2$.
Therefore,
\begin{equation}
    \Delta x_4 \leq \sqrt{\frac{\sqrt{3}}{2}\varepsilon\cdot 2} = \sqrt[4]{3}\,\sqrt{\varepsilon}.\label{Eq:DeltaX4L2}
\end{equation}
We claim that the bound is attained for $\vect{u}=\frac{1}{\sqrt{3}}\left[1,1,1\right]^\top$ on the equator:
Suppose $\vect{u}$ rounds inward to $\tilde{\vect{u}}=\vect{u}-\frac{\varepsilon}{2}\left[1,1,1\right]^\top$.
Then $\Vert\tilde{\vect{u}}\Vert_2^2=1-\sqrt{3}\,\varepsilon+\frac{3}{4}\varepsilon^2$, and
$\Delta x_4=\tilde{x}_4=\sqrt{\sqrt{3}\,\varepsilon-\frac{3}{4}\varepsilon^2}=\sqrt[4]{3}\,\sqrt{\varepsilon}+\mathcal{O}\left(\varepsilon^{3/2}\right)$,
which matches the bound in \cref{Eq:DeltaX4L2} up to $\mathcal{O}\left(\varepsilon^{3/2}\right)$.

Since $\Vert\vect{u}-\tilde{\vect{u}}\Vert_2\in\mathcal{O}\left(\varepsilon\right)$ and $\Delta x_4\in\mathcal{O}\left(\sqrt{\varepsilon}\right)$, the change of $x_4$ dominates, and $\Delta Q_\mathrm{exact}\leq\Delta x_4+\mathcal{O}\left(\varepsilon\right)$.
The chord between $\uvec{x}$ and $\tilde{\uvec{x}}$ is at least $\Delta x_4$, since $\Delta x_4$ is one of its components.
Further, the arc $\Delta Q_\mathrm{exact}$ is at least as long as the chord.
Hence, $\Delta x_4\leq\Delta Q_\mathrm{exact}\leq\Delta x_4+\mathcal{O}\left(\varepsilon\right)$.
As in \cref{Sec:ErrorMetrics}, $\Delta Q$ denotes the leading term.
Here, the remainder is $\mathcal{O}\left(\varepsilon\right)$.
We conclude
\begin{equation}
    \boxed{
        \begin{aligned}
            \Delta Q^{\lTwo,\mathrm{domain}} &= \sqrt[4]{3}\,\sqrt{\varepsilon},\\
            \vect{u}^{\lTwo,\mathrm{domain}} &=\frac{1}{\sqrt{3}}\left[1,1,1\right]^\top.
        \end{aligned}}
    \label{Eq:L2DomainResult}
\end{equation}

\subsection{Range Quantization}\label{Sec:LTwoRangeProof}
Let $\vect{u}$ be the parameters we wish to quantize and $s={1-\Vert \vect{u} \Vert_2^2}$.
By \cref{Eq:L2BoundX4}, a candidate $\tilde{\vect{u}}$ with $t=1-\Vert\tilde{\vect{u}}\Vert_2^2$ has $\Delta x_4\leq\sqrt{\varepsilon}$ if $\vert s-t\vert\leq\varepsilon$.
We show that such a candidate always exists.

Say $\vect{u}$ lies in a full cell and let $\tilde{\vect{u}}^{(0)}$ be its corner closest to the origin.
We walk to the opposite corner via $\tilde{\vect{u}}^{(1)}=\tilde{\vect{u}}^{(0)}+\varepsilon\vect{e}_1$, $\tilde{\vect{u}}^{(2)}=\tilde{\vect{u}}^{(1)}+\varepsilon\vect{e}_2$, and $\tilde{\vect{u}}^{(3)}=\tilde{\vect{u}}^{(2)}+\varepsilon\vect{e}_3$.
From one corner to the next, the squared norm increases by at most $2\varepsilon$, e.g., $\left(\tilde{u}^{(0)}_1+\varepsilon\right)^2-\left(\tilde{u}^{(0)}_1\right)^2=\varepsilon\left(2\tilde{u}^{(0)}_1+\varepsilon\right)\leq2\varepsilon$, since $\tilde{u}^{(0)}_1+\varepsilon\leq1$.
Since $\Vert\vect{u}\Vert_2^2$ lies between $\Vert\tilde{\vect{u}}^{(0)}\Vert_2^2$ and $\Vert\tilde{\vect{u}}^{(3)}\Vert_2^2$, one corner is within $\varepsilon$ of it, i.e., $\vert s-t\vert\leq\varepsilon$.

The walk needs a full cell, which exists everywhere except for $u_1>1-\frac{\varepsilon}{2}$, where the unit ball domain touches the boundary of $\left[-1,1\right]^3$ at $\vect{e}_1$.
There, every candidate has $\tilde{u}_1=1-\frac{\varepsilon}{2}$, so $s,t\leq1-\left(1-\frac{\varepsilon}{2}\right)^2\leq\varepsilon$, and again $\vert s-t\vert\leq\varepsilon$.
This bound is attained at $\vect{u}=\vect{e}_1$, where $s=0$ and every candidate has $t=\varepsilon+\mathcal{O}\left(\varepsilon^2\right)$, so $\Delta x_4=\sqrt{\varepsilon}+\mathcal{O}\left(\varepsilon^{3/2}\right)$.
With the chord argument of the domain case, we conclude
\begin{equation}
    \boxed{
        \begin{aligned}
            \Delta Q^{\lTwo,\mathrm{range}} &= \sqrt{\varepsilon},\\
            \vect{u}^{\lTwo,\mathrm{range}} &=\left[1,0,0\right]^\top.
        \end{aligned}}
\end{equation}

%% file: secs/S1symbols.tex
% The symbol tables cannot break, so do not stretch the space between them to fill columns.
\raggedbottom
\section{Symbols}
The tables list the symbols of the paper and of this \SupplementalName.
Symbols used by a single proof come last, with the section of that proof.

\input{tabs/ListOfSymbols.tex}

%% file: tabs/ListOfSymbols.tex
% Symbols of the paper and of the supplemental material, grouped by topic.
% Each group is a table of its own, so the list can break between columns and pages.
% All tables share the width of the symbol column, so their columns line up.
\noindent\begin{tblr}{colspec={Q[l,wd=16mm]X[l]}, width=\linewidth, rowsep=1pt}
    \toprule
    \SetCell[c=2]{l}\textbf{Notation} & \\
    \midrule
    $\vect{x}$, $x_i$                               & vector and its $i$th component\\
    $\uvec{x}$, $\ucomp{x}_i$                       & unit vector, $\Vert\uvec{x}\Vert_2=1$, and its $i$th component\\
    $\vect{x}_{1:3}$                                & first three components of $\vect{x}$\\
    $\quant{x}$, $\quant{\vect{x}}$                 & quantized scalar and vector\\
    $\aprx{x}$, $\aprx{\vect{x}}$                   & reconstructed scalar and vector\\
    $\vect{0}$, $\vect{1}$                          & vectors whose components are $0$ and $1$\\
    $\vect{e}_i$                                    & $i$th canonical unit vector\\
    $\mat{I}$                                       & identity matrix\\
    $\mathrm{norm}_k\left(\vect{x}\right)$          & normalization $\vect{x}/\Vert\vect{x}\Vert_k$\\
    $\mathrm{neg}\left(x\right)$                    & $1$ if $x<0$, else $0$\\
    $\mathrm{round}\left(x\right)$                  & nearest integer, per component\\
    $\angle\left(\uvec{x},\aprx{\uvec{x}}\right)$   & angle between $\uvec{x}$ and $\aprx{\uvec{x}}$\\
    $\mathbb{E}$                                    & expected value\\
    $S^2$, $S^3$                                    & unit sphere and unit hyper-sphere\\
    $n$                                             & number of components of a unit vector\\
    \bottomrule
\end{tblr}

\bigskip
\noindent\begin{tblr}{colspec={Q[l,wd=16mm]X[l]}, width=\linewidth, rowsep=1pt}
    \toprule
    \SetCell[c=2]{l}\textbf{Parameterizations} & \\
    \midrule
    $\vect{f}$                                      & parameterization, decompresses $\vect{u}$ and $\vect{c}$ to a unit vector; a subscript names the method\\
    $\vect{f}^{-1}$                                 & compression, maps a unit vector to $\vect{u}$ and $\vect{c}$\\
    $\vect{u}$, $u_i$                               & (domain) parameters\\
    $\vect{c}$, $c$, $c_1$, $c_2$                   & chart indices\\
    $C$, $C_i$                                      & number of charts, in total and for $c_i$\\
    $\vect{v}$, $v_i$                               & unnormalized vector, $\vect{f}=\mathrm{norm}_2\left(\vect{v}\right)$\\
    $\sigma$                                        & squared norm $\Vert\vect{v}\Vert_2^2$\\
    $i_\mathrm{max}$                                & index of the component of largest magnitude\\
    $\vect{P}_{i_\mathrm{max}}$                     & permutation, $\vect{P}_{i_\mathrm{max}}^{-1}$ moves component $i_\mathrm{max}$ to the end\\
    $w$, $\vect{w}$                                 & warp function, $\vect{w}$ applies $w$ to each component\\
    $\vect{u}'$                                     & warped parameters $\vect{w}\left(\vect{u}\right)$ of $\lInfT$\\
    $w\left(x;a\right)$                             & cubic warp $a\,x^3+\left(1-a\right)x$\\
    $w_\mathrm{tan}$                                & tangent warp $\tan\left(\frac{\pi}{4}x\right)$\\
    $a$, $a^\ast$                                   & cubic warp parameter, minimizer of $\kappa^\mathrm{range}$\\
    $a_\mathrm{tan}$                                & $a$ of the Remez fit of $w_\mathrm{tan}$\\
    $\tau$                                          & scale factor of $\vect{f}_\mathrm{Exp}^{-1}$\\
    $\upsilon$                                      & scaled norm $\frac{\pi}{2}\Vert\vect{u}\Vert_2$ of $\vect{f}_\mathrm{Exp}$\\
    \bottomrule
\end{tblr}

\bigskip
\noindent\begin{tblr}{colspec={Q[l,wd=16mm]X[l]}, width=\linewidth, rowsep=1pt}
    \toprule
    \SetCell[c=2]{l}\textbf{Parameterization methods} & \\
    \midrule
    $\lTwo$                                         & drops a fixed component\\
    $\lTwoI$                                        & drops the component of largest magnitude\\
    $\lOne$                                         & projects onto the hyper-octahedron\\
    $\lInf$                                         & projects onto the hyper-cube\\
    $\lInfT$                                        & $\lInf$ with a warp function $w$\\
    $\lInfTan$                                      & $\lInfT$ with $w_\mathrm{tan}$\\
    $\lInfCub$                                      & $\lInfT$ with $w\left(x;a^\ast\right)$\\
    $\mathrm{Exp}$                                  & rotation axis scaled by half the angle\\
    $\mathrm{Qu}$                                   & quantizes all four components\\
    \bottomrule
\end{tblr}

\bigskip
\noindent\begin{tblr}{colspec={Q[l,wd=16mm]X[l]}, width=\linewidth, rowsep=1pt}
    \toprule
    \SetCell[c=2]{l}\textbf{Error metrics} & \\
    \midrule
    $\vect{p}$                                      & input position in its cell, in units of $\varepsilon$\\
    $\vect{b}$                                      & chosen corner of the cell\\
    $\mathcal{B}$                                   & candidate corners, $\mathcal{B}^\mathrm{domain}$ or $\mathcal{B}^\mathrm{range}$\\
    $\vect{y}$                                      & rounding error $\vect{p}-\vect{b}$, in units of $\varepsilon$\\
    $\alpha$                                        & angle between the input and the chosen corner\\
    $\mat{J}_\vect{f}$, $\mat{J}_\vect{v}$          & Jacobian matrices of $\vect{f}$ and $\vect{v}$\\
    $\mat{G}$                                       & Gram matrix $\mat{J}_\vect{f}^\top\mat{J}_\vect{f}$\\
    $\mat{F}$                                       & projection onto the tangent space at $\vect{f}$, $\mat{I}-\vect{f}\vect{f}^\top$\\
    $Q$                                             & first-order squared error $\vect{y}^\top\mat{G}\,\vect{y}$, in units of $\varepsilon^2$\\
    $\vect{z}$                                      & rounding error carried over to $\vect{v}$, $\mat{J}_\vect{v}\,\vect{y}$\\
    \bottomrule
\end{tblr}

\bigskip
\noindent\begin{tblr}{colspec={Q[l,wd=16mm]X[l]}, width=\linewidth, rowsep=1pt}
    \toprule
    \SetCell[c=2]{l}\textbf{Quantization and errors} & \\
    \midrule
    $U$                                             & number of states per parameter\\
    $\varepsilon$                                   & sample spacing $\frac{2}{U}$\\
    $q$, $q^{\sim1}$                                & quantization and dequantization\\
    $k$                                             & number of bits of a compressed quaternion\\
    $N$                                             & number of codes, $2^k$\\
    $\Delta Q$                                      & maximum quantization error\\
    $\Delta Q^\mathrm{domain}$                      & $\Delta Q$ of domain quantization\\
    $\Delta Q^\mathrm{range}$                       & $\Delta Q$ of range quantization\\
    $\Delta Q^\mathrm{lower}$                       & lower bound on $\Delta Q$ for $N$ codes\\
    $\vect{u}^\mathrm{domain}$                      & location of $\Delta Q^\mathrm{domain}$\\
    $\vect{u}^\mathrm{range}$                       & location of $\Delta Q^\mathrm{range}$\\
    $\kappa$                                        & compactness factor in degrees, $\Delta Q\approx\kappa\cdot2^{-k/3}$\\
    $\kappa^\mathrm{domain}$                        & $\kappa$ of $\Delta Q^\mathrm{domain}$\\
    $\kappa^\mathrm{range}$                         & $\kappa$ of $\Delta Q^\mathrm{range}$\\
    $\kappa^\mathrm{lower}$                         & $\kappa$ of $\Delta Q^\mathrm{lower}$\\
    $e$                                             & color error $\max\left(\vert\Delta R\vert,\vert\Delta G\vert,\vert\Delta B\vert\right)$ of a pixel\\
    \bottomrule
\end{tblr}

\bigskip
\noindent\begin{tblr}{colspec={Q[l,wd=16mm]X[l]Q[r]}, width=\linewidth, rowsep=1pt}
    \toprule
    \SetCell[c=2]{l}\textbf{Symbols of single proofs} & & Section\\
    \midrule
    $\rho$                                          & $v_1^2+v_2^2+v_3^2$ & \ref{Sec:LInfTanDomainProof}\\
    $s$                                             & $\sin\left(\upsilon\right)/\Vert\vect{u}\Vert_2$ & \ref{Sec:ExpDomainProof}\\
    $\uvec{u}$                                      & direction $\vect{u}/\Vert\vect{u}\Vert_2$ & \ref{Sec:ExpDomainProof}\\
    $P$, $m$                                        & $\vect{1}^\top\vect{p}$ and $\mathrm{round}\left(P\right)$ & \ref{Sec:LOneRangeProof}\\
    $\pmb{\omega}$                                  & weights $\vect{u}/v_4$ & \ref{Sec:LTwoIRangeProof}\\
    $\xi$                                           & random threshold, uniform in $[0,1)$ & \ref{Sec:LTwoIRangeProof}\\
    $s_k$, $T_k$                                    & running sums of $\vect{p}$ and $\vect{b}$ & \ref{Sec:LTwoIRangeProof}\\
    $r_k$                                           & fractional part of $s_k$ & \ref{Sec:LTwoIRangeProof}\\
    $d_k$                                           & error $s_k-T_k$ of a running sum & \ref{Sec:LTwoIRangeProof}\\
    $\lambda_k$                                     & convex weights from $\pmb{\omega}$ & \ref{Sec:LTwoIRangeProof}\\
    $x_4$, $\aprx{x}_4$                             & fourth component of $\lTwo$ and its reconstruction & \ref{Sec:LTwoProof}\\
    $\Delta x_4$                                    & $\vert\aprx{x}_4-x_4\vert$ & \ref{Sec:LTwoProof}\\
    $s$, $t$                                        & $1-\Vert\vect{u}\Vert_2^2$ and $1-\Vert\aprx{\vect{u}}\Vert_2^2$ & \ref{Sec:LTwoProof}\\
    $\aprx{\vect{u}}^{(j)}$                         & corners on a walk through the cell & \ref{Sec:LTwoRangeProof}\\
    \bottomrule
\end{tblr}